\documentclass[12pt]{article}
\pdfoutput=1
\usepackage{latexsym,epsfig,amssymb,amsmath,amsthm,color,url,bbm,pdfsync,tikz}
\usetikzlibrary{arrows,positioning} 
\usepackage{graphicx}
\usepackage{authblk}
\usepackage{inputenc}
\usepackage[square,comma,sort&compress]{natbib}
\setcitestyle{numbers,multirow,import}
\usepackage{graphicx}
\allowdisplaybreaks
\numberwithin{equation}{section}
\numberwithin{figure}{section}
\numberwithin{table}{section}

\newtheorem{Theorem}{Theorem}[section]
\newtheorem{Lemma}[Theorem]{Lemma}
\newtheorem{Remark}[Theorem]{Remark}
\newtheorem{Example}[Theorem]{Example}
\newtheorem{Proposition}[Theorem]{Proposition}
\newtheorem{Definition}[Theorem]{Definition}
\newtheorem{Corollary}[Theorem]{Corollary}

\newcommand{\bthe}{\begin{Theorem}}
\newcommand{\ethe}{\end{Theorem}}

\newcommand{\ble}{\begin{Lemma}}
\newcommand{\ele}{\end{Lemma}}

\newcommand{\bde}{\begin{Definition}}
\newcommand{\ede}{\end{Definition}}

\newcommand{\bco}{\begin{Corollary}}
\newcommand{\eco}{\end{Corollary}}

\newcommand{\bpr}{\begin{Proposition}}
\newcommand{\epr}{\end{Proposition}}

\newcommand{\brem}{\begin{Remark}}
\newcommand{\erem}{\end{Remark}}

\newcommand{\bexam}{\begin{Example}}
\newcommand{\eexam}{\end{Example}}

\newcommand{\beqq}{\begin{equation}}
\newcommand{\eeqq}{\end{equation}}

\newcommand{\beao}{\begin{eqnarray*}}
\newcommand{\eeao}{\end{eqnarray*}\noindent}

\newcommand{\beam}{\begin{eqnarray}}
\newcommand{\eeam}{\end{eqnarray}\noindent}

\newcommand{\barr}{\begin{array}}
\newcommand{\earr}{\end{array}}

\newcommand{\bproof}{\begin{proof}}
\newcommand{\eproof}{\end{proof}}

\newcommand{\dd}{\mathrm{d}}
\newcommand{\PP}{\textbf{P}}
\newcommand{\EE}{\textbf{E}}

\newcommand{\RR}{\mathbb{R}}

\begin{document}
\bibliographystyle{plain}

\title{Common Growth Patterns for Regional Social Networks: a Point Process Approach}

\author{Tiandong Wang \thanks{Department of Statistics, Texas A\&M University, twang@stat.tamu.edu}
and
Sidney I. Resnick \thanks{School of Operations Research and Information Engineering, Cornell University,sir1@cornell.edu}}

\maketitle

\begin{abstract}
  Although recent research on social networks emphasizes microscopic
  dynamics such as retweets and social connectivity of an individual
  user, we focus on macroscopic growth dynamics of social network link
  formation.  Rather than focusing on one particular dataset, we find
  invariant behavior in regional social networks that are
  geographically concentrated.  Empirical findings suggest that
  the startup phase of a regional network can be modeled by a
  self-exciting point process. After the startup phase ends, the
  growth of the links can be modeled by a non-homogeneous Poisson
  process with constant rate across the day but varying rates from day
  to day, plus a nightly inactive period when local users are expected
  to be asleep. Conclusions are drawn based on analyzing four
  different datasets, three of which are regional and a
  non-regional one is included for contrast.
\end{abstract}

\section{Introduction}
Network growth can be measured in several ways. A large number of
links or edges can indicate that users find the network useful and
compelling and may encourage potential users to join.  A large number
of nodes may be a driver of network revenue from subscriptions or
advertising revenue. Based on understanding growth patterns, network companies can
decide how and when to conduct stimuli that maintain and enhance network
attractiveness and usefulness.
In this paper we focus on statistically describing the process
of edge formation for regionally restricted social networks.

Social network analysis has triggered an exploding number of research
topics.  In the network and computer science literature, many novel
models based on point processes have been proposed to analyze network
evolution.  For example, in \cite{zhao:2015}, a self-exciting point
process (SEPP) is used to predict the number of retweets for a given
post on Twitter. Subsequently data-driven refinements of the SEPP
model were proposed \cite{kobayashi:2016, rizoiu2017expecting,
  mishra2016feature,martin2016exploring, srijith2017longitudinal}.
often emphasizing user based microscopic network evolution, e.g. the cascading of
a given tweet and individual's social connectivity dynamics
(\cite{zang2017long}).  There has been less focus on the macroscopic
growth of the whole network.

Existing network growth models include preferential attachment models
\cite{bollobas:borgs:chayes:riordan:2003, krapivsky:2001,
  krapivsky:redner:2001}, the Bass model \cite{mahajan1990new} and the
Susceptible-Infected model \cite{anderson1992infectious} but these
models do not consider link growth in real time. Different
models have recently been proposed for example in \cite{zang2016beyond,
  mislove2008growth}, with primary focus on modeling the growth of a
particular network, rather than looking for invariants across
 networks.
We emphasize common
properties across regional networks that are
largely localized to one time zone.  We focus on the growth of edges
in the network, which is a measure of the utility network users find
in a network.

In a temporal network, link creation is often accompanied by
timestamps and  a natural way to study  network growth is to use
the timestamps.  In a graphical
representation of social networks, edges often represent individual
interactions between two users.  Some of these networks are regional
in that users in the network are geographically concentrated
(e.g. comments left on the talk page of articles in Dutch Wikipedia),
compared to less geographically concentrated ones (e.g. comments left on the talk
page of articles in English Wikipedia).  Using a point process
approach, we observe some common patterns of network growth in regional
social networks.

\subsection{Summary of findings}
When an edge is created at a certain timestamp, we regard it as a point or event observed in a point process.
At first, the network evolution can be modeled by a SEPP with an exponential triggering function.
After the startup phase of the network ends, the point process is close to a non-homogeneous Poisson process (NHPP), 
apart from a nightly inactive period when most users in the same regional locale are asleep at the same time and
hence not active on the network.
We discuss four different network datasets listed on KONECT (\url{konect.cc/networks}):
\begin{itemize}
\item Facebook wall posts for users in New Orleans.
\item Dutch Wikipedia Talk.
\item German Wikipedia Talk.
\item MathOverflow.
\end{itemize}
The first three networks are regional but the last one is not and is included for contrast.

The rest of the paper is organized as follows. In Section~\ref{sec:prelim}, we summarize background knowledge on a SEPP
 and outline an estimation algorithm (cf. \cite{kirchner:2017, kirchner:2018}) which is used throughout the paper when estimating
 parameters in a SEPP. Analyses on the four network datasets are included in Section~\ref{sec:data} and 
 Section~\ref{sec:discuss} highlights key findings in this study and raises open problems for future study.

\section{Preliminaries}\label{sec:prelim}
In this section, we include preliminary knowledge for point processes.
\subsection{Self exciting point processes (SEPP)}\label{subsec:sepp}
One of the initial use of the SEPP is to model earthquake occurrences \cite{ogata1988statistical}, but nowadays the SEPP model
has a variety of applications, e.g. modeling financial data
\cite{bacry2015hawkes, hawkes:2018, embrechts2011multivariate}, social
media \cite{zhao:2015, rizoiu:2017,
  rizoiu2017expecting,srijith2017longitudinal}, terrorist activity
\cite{porter2012self}, as well as crime and security
\cite{mohler2013modeling}.  Here is a brief overview.

A univariate point process is constructed on a probability space $(\Omega, \PP,\mathcal{F})$ where 
points are viewed as events.
Define on this probability space a sequence of stochastic event times $0= T_0\le T_1\le T_2\le\ldots\le T_n\le T_{n+1}\le\ldots$, and
using the standard point measure notation
\[
\epsilon_x(A) := 
\begin{cases}
1,&\qquad \text{if $x\in A$},\\
0,&\qquad \text{if $x\notin A$}.
\end{cases}
\]
 the counting process on the Borel sets of $\RR_+$, $\mathcal{B}(\RR_+)$, is defined as
\[
N(\cdot) := \sum_{i = 1}^\infty \varepsilon_{T_i}(\cdot).
\]
We denote $\mathcal{H}_t$ as the \emph{history} associated with $N$.

A \emph{self-exciting} point process (SEPP) is a point process which has
a conditional intensity (cf. \cite{daley:vere-jones:2003}) consisting of
\begin{itemize}
\item  A \emph{baseline or background intensity} $\eta\ge 0$ and
\item A \emph{triggering function} $g:\RR_+\mapsto \RR_+$ measuring how much a recent event increases the intensity.
\end{itemize}
Mathematically, the conditional intensity is of the following assumed form:
\beqq\label{eq:cond_intens}
\lambda(t|\mathcal{H}_t) = \eta + K\int_0^t g(t-s)\dd N(s),\qquad \eta,K >0, t\ge 0.
\eeqq
For brevity we often suppress explicitly writing the history
$\mathcal{H}_t$ and simply use $\lambda(t)$ to denote the conditional intensity.
The point process $N(\cdot)$ is also known as the
\emph{Hawkes process} in the literature.

There are two common choices of the triggering function $g$, namely
\begin{itemize}
\item Exponential function: 
\beqq\label{eq:exp_sepp}
g(t; \beta) = \beta e^{-\beta t}, \qquad\text{for some } \beta>0, t\ge 0;
\eeqq
\item Power-law function: 
\beqq\label{eq:pl_sepp}
g(t; c, \gamma) = \frac{\gamma-1}{(1+t)^{\gamma}}, \qquad\text{for some } \gamma>1, t\ge 0;
\eeqq
\end{itemize}
and  we only focus on these  two choices.

We later outline an estimation procedure in Section~\ref{subsec:est_sepp} for a SEPP based on \cite{kirchner:2018}, which 
approximates a SEPP by an integer-valued autoregressive model with infinite order, $\text{INAR}(\infty)$. 
Theoretical justifications of the estimation procedure are given in \cite{kirchner:2016},
where the stationarity of both the SEPP and $\text{INAR}(\infty)$  is assumed.
In fact, the stationarity of the SEPP is 
obtained from a branching representation as given in
\cite{hawkes1974}.
Here with triggering functions specified in either \eqref{eq:exp_sepp} or \eqref{eq:pl_sepp},
the SEPP is stationary if $K<1$.

\subsection{Estimation methods for a SEPP}\label{subsec:est_sepp}
In terms of estimation, one approach is to use maximum likelihood estimation (MLE). 
Let $\boldsymbol{\theta}$ be parameters in a SEPP, $N$,
so $\boldsymbol{\theta} = (\eta, K,\beta)$ if the triggering function is exponential as in \eqref{eq:exp_sepp} and
$\boldsymbol{\theta} = (\eta, K,\gamma)$ if the triggering function is power-law as in \eqref{eq:pl_sepp}.
Write $N(t) := N((0,t])$ and $0= T_1\le T_2\le\ldots\le T_n$ are $n$ event times before time $t$ where $N(t) = n$.
The log-likelihood function for $N$ based on observing $\{T_i\}_{i=1}^n$ (cf. \cite{daley:vere-jones:2003})  is
\beqq\label{eq:loglik}
l\Bigl(\boldsymbol{\theta};\{T_i\}_{i=1}^n\Bigr) = -\int_0^t \lambda(s)\dd s + \sum_{i=1}^{N(t)} \log(\lambda(T_i)),
\eeqq
and the conditional intensity function $\lambda$ is calculated as in
\eqref{eq:cond_intens} with triggering function $g$ that
is either exponential \eqref{eq:exp_sepp} or power-law \eqref{eq:pl_sepp}.
Then the MLE is calculated as
\[
\hat{\boldsymbol{\theta}}_\text{MLE} := \text{argmax}_{\boldsymbol{\theta}} \Bigl(\boldsymbol{\theta};\{T_i\}_{i=1}^n\Bigr),
\]
where the optimization can be done by the R function \verb6optim6 using methods such as \verb6L-BFGS-B6.
However, evaluating \eqref{eq:loglik} takes $O(n^2)$ time and the log-likelihood can be almost flat 
in a large region of the parameter space; see \cite{veen:2008} for details.
When applied to real data, the numerical optimization results are also sensitive to the choice of initial values of the model parameters.
In \cite{kirchner:2017, kirchner:2018}, a faster estimation method is introduced. The main idea is to discretize time and 
approximate the count process on the resulting intervals by an integer-valued autoregressive model with finite order $p$, i.e. $\text{INAR}(p)$. 
We now outline this estimation procedure.

Consider a SEPP $N$ on a time window $[0,t]$.
Choose a small $\Delta$ such that $0<\Delta<<t$ and some
$p\in\mathbb{N}$ such that $p\Delta$ 
is large. This decomposes $[0,t]$
into roughly $t/\Delta$ bins.
Here we do not choose too small a $\Delta$ as it will create too many zeros in the bin counts and therefore
makes the estimation based on INAR($p$) less reliable.
See \cite{kirchner:2017} for a discussion on the choice of $p$ and $\Delta$.
\begin{enumerate}
\item Discretization: calculate the number of edges added at times
  falling within each
  bin of length $\Delta$ to get the bin-count sequence $X_n^{(\Delta)}$, i.e.
\[
X_n^{(\Delta)}: = N\bigl(((n-1)\Delta, n\Delta]\bigr),\quad n=1,2,\ldots, [t/\Delta].
\]
\item Optimization: find conditional LS estimates, i.e.
\beqq\label{eq:opt}
\left(\hat{\alpha}^{(\Delta)}_0,\ldots, \hat{\alpha}^{(\Delta)}_p\right):=
\text{argmin}_{(\alpha_0,\ldots,\alpha_p)\in\mathbb{R}^{p+1}}\sum_{n=p+1}^{[t/\Delta]}\left(X_n^{(\Delta)}-\alpha_0-\sum_{k=1}^p\alpha_k X_{n-k}^{(\Delta)}\right)^2.
\eeqq
\item Normalization:  Set,
\[
\hat{\eta} := \frac{\hat{\alpha}^{(\Delta)}_0}{\Delta},\quad \hat{K}:=\sum_{k=1}^p\hat{\alpha}^{(\Delta)}_k,
\quad \hat{g}_k^{(\Delta)} := \frac{\hat{\alpha}^{(\Delta)}_k}{\hat{K}\Delta}.
\]
\end{enumerate}
Note that $ \hat{g}_k^{(\Delta)}$ is an estimator for $g(k\Delta)$.
We assume the triggering function $g$ belongs to one of the two
parametric density families $(g_{\theta'})_{\theta'\in\Theta'}$ 
 given in \eqref{eq:exp_sepp} and \eqref{eq:pl_sepp} and 
 the unknown 
parameters are estimated from 
\beqq\label{eq:NLS}
\hat{\boldsymbol\theta} :=\text{argmin}_{\boldsymbol{\theta}'\in\Theta'} \sum_{k=1}^p \Bigl(g\bigl((k-0.5\bigr)\Delta;\boldsymbol{\theta}')-
\hat{g}_k^{(\Delta)}\Bigr)^2.
\eeqq
We use the R function \verb6nls6 using an algorithm from the \verb6port6 library to solve the optimization problem in \eqref{eq:NLS}.

The procedure above actually assumes the  approximation,
\[
\EE(X_n^{(\Delta)}|\mathcal{H}_{(n-1)\Delta}) \approx \eta\Delta +
K\Delta \int_{(n-p-1)\Delta}^{(n-1)\Delta} g(n\Delta - u)\dd N(u)\]
and thus only observations within $((n-p-1)\Delta, (n-1)\Delta]$ have
impacts on the triggering function $g(\cdot)$. The SEPP, therefore,
has finite memory; see \cite{jo2015correlated, zang2017long} for more detail.

When applying the estimation procedure to real datasets in Section~\ref{sec:data},
we divide all timestamps into non-overlapping time windows ({not
  to be confused with the bins of length $\Delta$ discussed above})
  with equal
number of events and fit the SEPP model to each time window, rather
than 
the whole dataset. {While this exposes us to the dangers of
overfitting, the reasons for this action are} three-fold:
\begin{enumerate}
\item As the network evolves, there might be some change points in the evolution so that modeling the whole network by one simple model
is not appropriate.
\item As noted in \cite{kirchner:2017}, if too many data points
are included, then it is more likely to have abnormal observations which are not captured by the underlying model. 
Therefore, formal statistical tests will likely reject the model when assessing the goodness of fit for the SEPP model.
\item At the beginning of the network, few events are observed, so it takes a long time to accumulate a certain number of points 
upon which we fit the SEPP model. Hence, we set up the time windows by
accumulating enough events, not by  using 
a fixed period of time.
\end{enumerate}

\section{Network Data}\label{sec:data}
\subsection{Facebook wall posts.}
The Facebook wall post dataset is data from a regional network of
users in New Orleans from 2004-09-13 to 2009-01-21.  The data forms a
directed graph where nodes are Facebook users and each directed edge
represents a post from one node to another node's page. The dataset
has three columns and 876,993 rows.  The first two columns are
anonymized user identifiers and the third column is a UNIX timestamp
marking the time of the wall post.  For each row, the second user
posts on the first user's wall at the time given in the third column.

\begin{figure}[h]
\centering
\includegraphics[scale=.5]{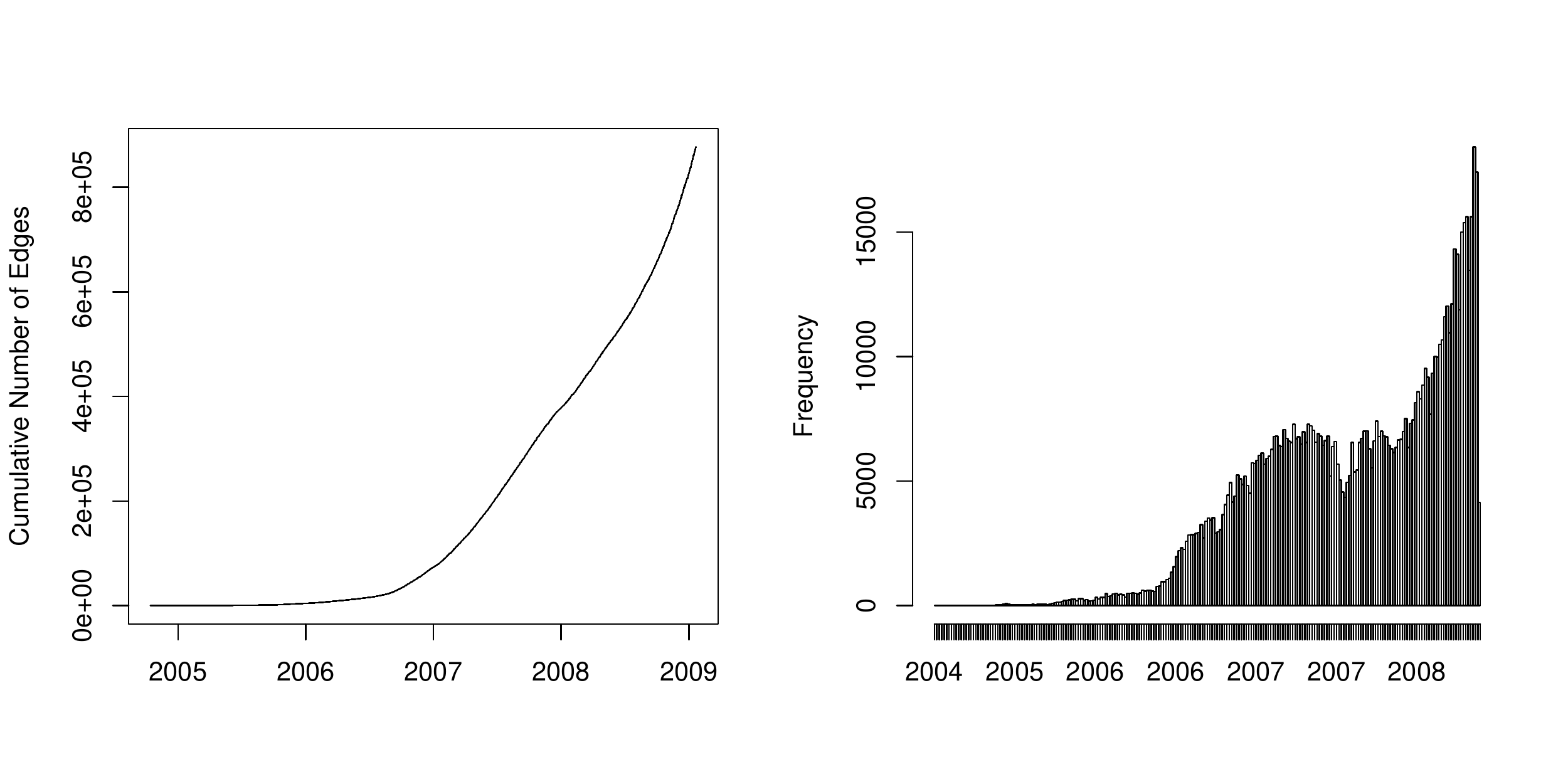}
\caption{Facebook wall posts  for users in New Orleans. Left:
  Cumulative number of edges 
  over time. Right: 
  number of new edges added per week.}\label{fig:fb_evo} 
\end{figure}
We plot the evolution of this regional network in Figure~\ref{fig:fb_evo}.
The left graph gives cumulative number of edges (posts) in the network over time and the right panel displays the total number of new
edges added per week. 
Both plots show that the network grows slowly during the first couple
of years and after year 2006 the growth pattern becomes relatively
stable. There is a sudden increase starting from mid-2008 is observed
and the
authors of \cite{viswanath2009evolution} speculate this is a
consequence of Facebook’s new site design unveiled on 
July 20, 2008 which allows users to  directly view wall posts
through friends' feeds.

\begin{figure}[h]
\includegraphics[width = 10cm, height=5cm]{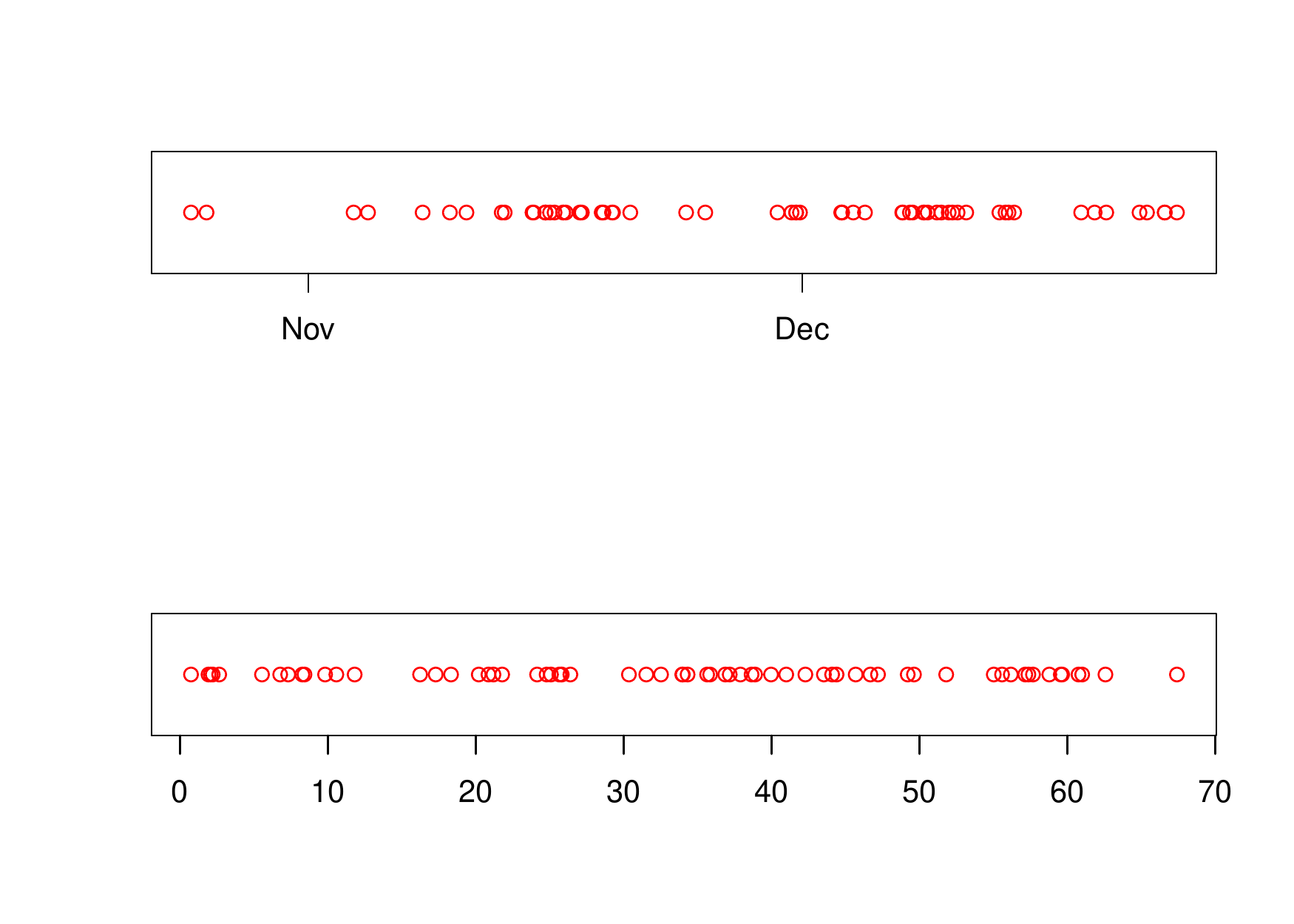}
\caption{Events observed from the Facebook wall post data from 2004-10-24 to 2004-12-23. (top) and a homogeneous Poisson process with unit rate (bottom).
In both cases, 65 events are included and the Facebook points are more bursty compared with Poisson points.}\label{fig:fb_pp}
\end{figure}
Based on plots in Figure~\ref{fig:fb_evo}, we start our analysis with
the Facebook data before 2006-01-01. 
The point process prior to 2006-01-01 is more bursty than after, i.e. more clustered in time, and later we justify this by calculating 
the burstiness parameter $B$ as defined below in \eqref{eq:burst}.
For graphical illustration, in the top panel of
Figure~\ref{fig:fb_pp}, we plot the 65 wall post time stamps occurring between
2004-10-24 and 2004-12-23. 
 For comparison, in the bottom panel of Figure~\ref{fig:fb_pp},
we plot 65 events generated from a homogeneous Poisson process with unit rate.
We see from Figure~\ref{fig:fb_pp} that the edge creation times
for Facebook New Orleans tends to be bursty.

One way to measure the burstiness of a point process \cite{goh2008burstiness} is through the \emph{burstiness parameter}, $B$, defined as
\beqq\label{eq:burst}
B := \frac{\sigma-\bar{\tau}}{\sigma+\bar{\tau}},
\eeqq
where $\sigma$ and $\bar{\tau}$ are the standard deviation and mean of the inter-event times (IETs), respectively.
In general, $B\in [-1,1]$ and a highly bursty point process will have
{$B$ close to $1$ since $\sigma$ is large relative to $\bar \tau$.}
For contrast, note that  a homogeneous Poisson process has $B=0$.

Calculating the burstiness parameter using all points prior to 2006-01-01 gives 
$B = 0.672$. The burstiness parameters for data after 2006 are summarized in the first row of Table~\ref{table:fb_burst}
and the estimates are less than 0.672.
The enhanced burstiness prior to 2006 suggests modeling the growth of the Facebook wall posts network by a SEPP,
but we need to determine the proper triggering function.
Since we only focus on the two choices in \eqref{eq:exp_sepp} and \eqref{eq:pl_sepp},
we make the decision by assessing the goodness of fit under the two triggering functions.

 
To fit the SEPP, we convert the time scale from seconds to days and 
apply the estimation procedure in Section~\ref{subsec:est_sepp} \cite{kirchner:2018}. We
set $p=150$, $\Delta = (1/24)/12$, which is a 5-minute time window, and fit the model locally to 55 disjoint time intervals each of which
contains 200 events. Note that no observations from the previous interval is used while fitting the model.
 This covers the time period from 2005-01-01 to 2006-05-01.
Due to the small number of observations at the beginning of the Facebook network, we do not further decrease the value of $\Delta$ since 
it will generate too many zeroes, making the optimization setup in \eqref{eq:opt} less accurate. 
Also, choosing too large a $p$ leads to error codes when doing computations in R so we do not further increase $p$.

Assuming a power law triggering function  \eqref{eq:pl_sepp}, we  apply the method outlined
in Section~\ref{subsec:est_sepp} and yielding the estimates 
$$\tilde{\boldsymbol{\theta}}
:= \Bigl(\tilde{\eta}, \tilde{K}, \tilde{\gamma}\Bigr).$$ To assess
the goodness of fit, we transform the event times
\beqq\label{eq:trans_time} \Lambda(T_i;\boldsymbol{\theta}) :=
\int_0^{T_i} \lambda(t|\mathcal{H}_t; \boldsymbol{\theta})\dd t,\qquad
i=1,\ldots, n, \eeqq and evaluate $\{\Lambda(T_i):i=1,\ldots,n\}$
using parameters estimated by $\tilde{\boldsymbol{\theta}}$ and with a
power-law triggering function, estimated values of the transformed
event times are \beqq\label{eq:trans_time_pl} {\Lambda}(T_i;
\tilde{\boldsymbol{\theta}}) := \tilde{\eta}\,T_i
+\tilde{K}\sum_{k=1}^{i-1}
\Bigl(1-(1+T_i-T_k)^{-(\tilde\gamma-1)}\Bigr) \eeqq If the model fit
is good, the transformed inter-event times, \beqq\label{eq:trans_iet}
\{\Lambda(T_i;\tilde{\boldsymbol{\theta}})-\Lambda(T_{i-1};\tilde{\boldsymbol{\theta}}):
i=1,\ldots, n\} \eeqq should be iid exponentially distributed with
unit rate (cf. \cite[Proposition 7.4.IV]{daley:vere-jones:2003}).
Examining the ACF plots (not included here) shows little
autocorrelation among transformed IETs.  Then we applied a
Kolmogorov-Smirnov (KS) test to see whether the transformed IETs
calculated as in \eqref{eq:trans_time_pl} and \eqref{eq:trans_iet})
follow an exponential distribution with unit rate.  It turns out that
the KS test does not reject the SEPP model with a power-law triggering
function \eqref{eq:pl_sepp} for only 1 out of the 55 disjoint time
windows, no matter whether the significance level, $\alpha$, is set to
0.01 or 0.05.  Moreover, estimates of the index of the power-law
function, $\gamma$, are large (with a median $\approx$ 28.6 based on
the 55 estimates), which indicates that the decay rate of the
triggering function is fast, and the poor acceptance rate by the KS
test suggests trying to fit the SEPP using an exponential triggering
function as in \eqref{eq:exp_sepp}.

Therefore, we refit the SEPP model to the same 55 disjoint time
intervals but this time used the exponential triggering
function. Setting $p=150$, $\Delta = (1/24)/12$ and applying the
method outlined in Section~\ref{subsec:est_sepp} give a new set of
estimates: \beqq\label{eq:nls} \widehat{\boldsymbol{\theta}} :=
\Bigl(\widehat{\eta}, \widehat{K}, \widehat{\beta}\Bigr).\eeqq Again,
we calculate the transformed times in \eqref{eq:trans_time} using
$\widehat{\boldsymbol{\theta}}$ and test the empirical distribution of
the transformed IETs against the hypothesis of an exponential
distribution with unit rate.  With an exponential triggering function
\eqref{eq:exp_sepp}, the transformed event times defined in
\eqref{eq:trans_time} become \beqq\label{eq:trans_time_exp}
{\Lambda}(T_i; \widehat{\boldsymbol{\theta}}) := \widehat{\eta}\,T_i
+\widehat{K}\sum_{k=1}^{i-1} \left(1-
  e^{-\widehat{\beta}(T_i-T_k)}\right), \qquad i=1,\ldots,n, \eeqq and
the transformed IETs are \beqq\label{eq:trans_iet_exp}
\{\Lambda(T_i;\widehat{\boldsymbol{\theta}})-\Lambda(T_{i-1};\widehat{\boldsymbol{\theta}}):
i=1,\ldots, n\}.  \eeqq From the ACF plots (not reported here), we do
not observe large autocorrelation among transformed IETs and the
acceptance rate of the KS test increases to 48.1\% with
$\alpha = 0.05$. The rate reaches 64.8\% when $\alpha = 0.01$.
Therefore, we conclude that the SEPP with an exponential triggering
function provides a better fit to the Facebook data from 2005-01-01 to
2006-05-01.  Note that due to the small number of events at the
beginning of the network, it takes a long time to accumulate 200
consecutive events.  In fact, the actual durations of the first 10
time windows are all longer than a week. For a long time duration, it is
possible to observe abnormal IETs which leads to rejections by the KS
test. After accounting for this fact, we think the 64.8\% acceptance
rate indicates a good fit of the SEPP model.

The analysis is then extended to the next 100 disjoint time windows each of which includes 200 points and this covers 
the time period from 2006-05-01 to 2006-09-22. However, 
the estimation algorithm in Section~\ref{subsec:est_sepp} fails when applied to data
subsequent to the 57-th time window, indicating a change
in the underlying model should be made. Looking at the the first 56 time windows (from 2006-05-01 to 2006-08-20), we see that the acceptance rate 
of the KS test is 40\% with $\alpha =0.05$ and 56.4\% with $\alpha =
0.01$.
{So  2006-05-01 to 2006-09-22 is a period where a model change seems to
take place.}

\begin{figure}[h]
\centering
\includegraphics[scale=.6]{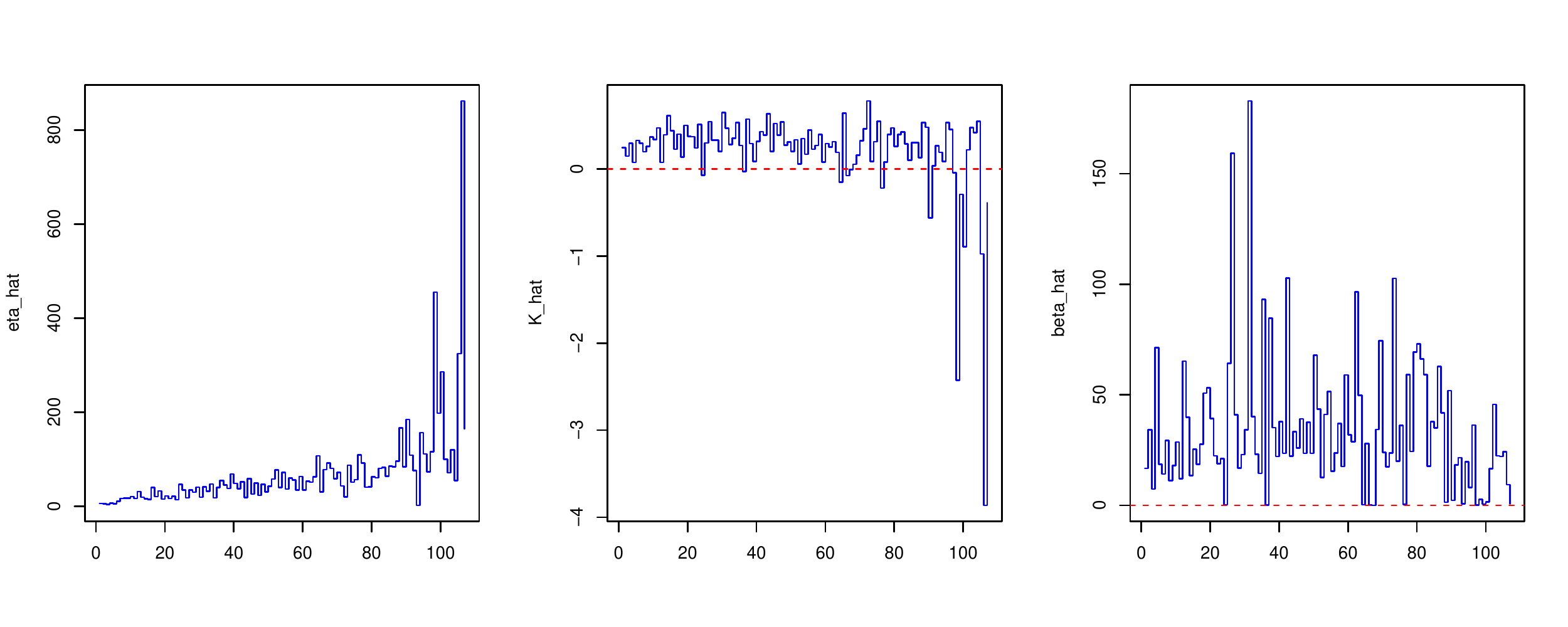}
\caption{Estimated values of $\widehat{\boldsymbol{\theta}} =
  \Bigl(\widehat{\eta}, \widehat{K}, \widehat{\beta}\Bigr)$ for the
  Facebook data from 
2005-01-01 to 2006-08-20. Each set of parameter estimates is
calculated using 200 points. The model struggles to fit after the 64-th time window.
}\label{fig:fb_est}
\end{figure}
Figure~\ref{fig:fb_est} summarizes the parameter estimates $\widehat{\boldsymbol{\theta}}$ (cf. \eqref{eq:nls}) over the total 110 disjoint time windows covering the period
from 2005-01-01 to 2006-08-20. Within each time window, 200 points are included. 
We notice that after the 64-th time window, $\widehat{K}$ starts to take negative values, which deviates from the model assumption
in \eqref{eq:exp_sepp}. 
Moreover, after the 64-th time window, negative values of $\widehat{K}$ are offset by large positive values of $\widehat{\eta}$.
For the two negative incidences of $\widehat{K}$ before the 64-th time window, 
the magnitude of $\widehat{K}$ is small (very close to 0) and
the KS test does not accept the model. Therefore, we conclude that changing the underlying model is necessary
from the 64-th time window onward (i.e. after 2006-05-24) and the SEPP with an exponential triggering function can only be used 
to model the Facebook data from 2005-01-01 to 2006-05-23.

\begin{figure}[h]
\centering
\includegraphics[scale=.5]{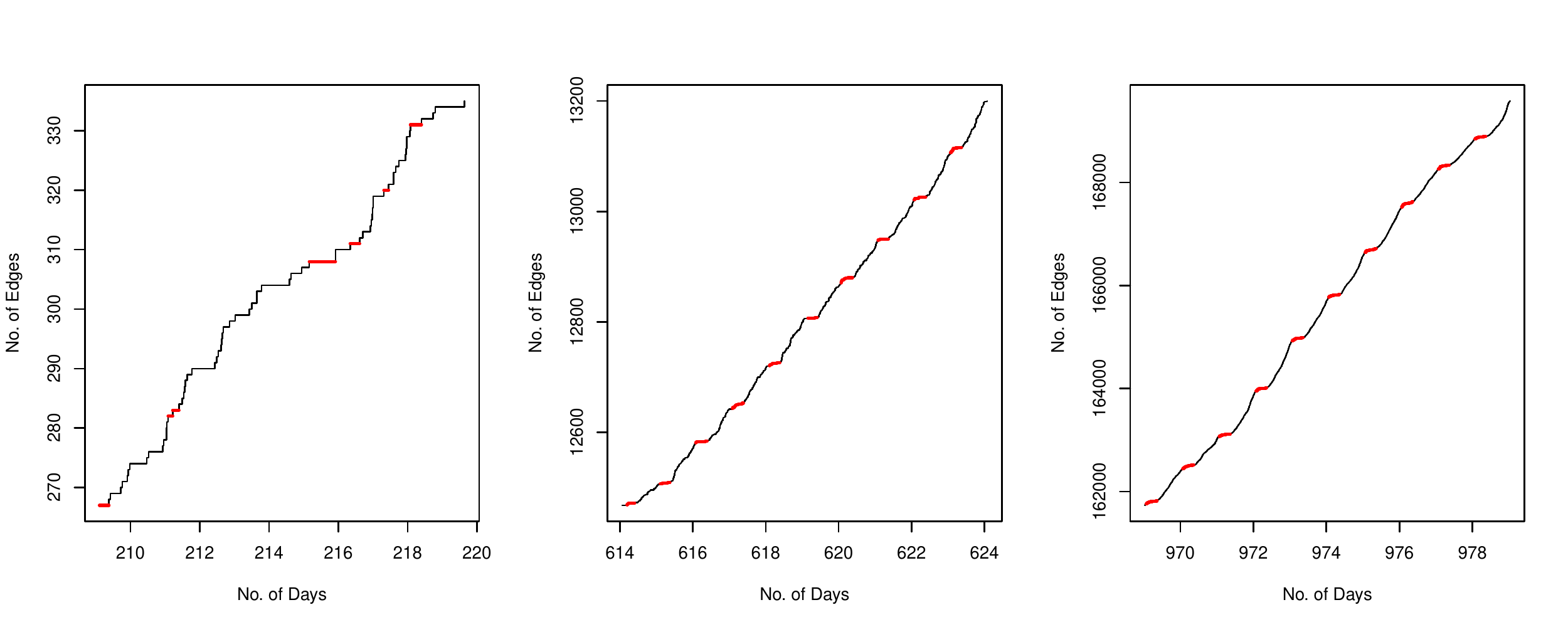}
\caption{Cumulative number of edges in two periods: (1) Left: 2005-04-11 to 2005-04-20 (when the SEPP model fits well);
(2) Middle: 2006-05-21 to 2006-05-30 (when the SEPP model fits poorly). (3) Right: 2007-05-11 to 2007-05-20.
 Red line segments refer to edges created during 1-8 AM each day (US Central Time Zone), when we expect sleep interferes with Facebook activity. 
The left plot looks more bursty and erratic than the other two and 
 a repetitive growth pattern is observed in the middle and right plots.}\label{fig:rep_fb}
\end{figure}
When inspecting the network growth pattern after 2006-05-24, we see a repeating pattern. For comparison,
in Figure~\ref{fig:rep_fb} we plot the cumulative number of edges from 3 representative periods:
 \begin{enumerate}
 \item 2005-04-11 to 2005-04-20, where the SEPP model fits well.
 \item 2006-05-21 to 2006-05-30, where the SEPP model starts to give poor fit to the data indicating a change in the model.
 \item  2007-05-11 to 2007-05-20, where a common repetitive pattern is observed.
 \end{enumerate}
The line segments in red correspond to the accumulation of new edges from 1 to 8 AM each day, during which we expect sleep interferes with Facebook activity, and the timestamps are adjusted 
according to the US Central Time Zone. 

\begin{table}[ht]
\begin{center}
\begin{tabular}{ccccc}
\hline
& Yr 2005 & Yr 2006 & Yr 2007 & Yr 2008  \\ 
\hline
$B_\text{all}$ & 0.405 & 0.454 & 0.386 & 0.370\\ 
$B_\text{active}$ & 0.421 & 0.407  & 0.112 & 0.111 \\ 
\hline
\end{tabular}
\end{center}
\smallskip
\caption{Burstiness parameters for IETs from 2005 to 2008, where $B_{all}$ is 
calculated using all IETs and $B_{active}$ is obtained by excluding all IETs that happen during 1-8 AM on each day.
Since 2006, we see a decrease in the burstiness parameter after removing IETs during users' inactive hours.}
\label{table:fb_burst}
\end{table}

In a plot of cumulative number of edges,  a relatively long flat line
segment  corresponds to a relatively long IET, meaning a
long period of inactivity in the network. 
Comparing the three panels in Figure~\ref{fig:rep_fb} shows that at first when the SEPP model is accepted, the longer IETs are
not necessarily associated with anticipated sleeping periods but later on
longer IETs are usually associated with the expected inactive hours (during 1-8 AM).
Moreover, Table~\ref{table:fb_burst} summarizes the burstiness parameter based on the point process from 2005 to 2008,
where $B_\text{all}$ is calculated using all timestamps during the year and $B_\text{active}$ excludes timestamps that occur during 1-8 AM each day.
Note that from 2006 onward, excluding timestamps during expected sleeping periods makes the point process less bursty.

\begin{figure}[h]
\centering
\includegraphics[scale=.5]{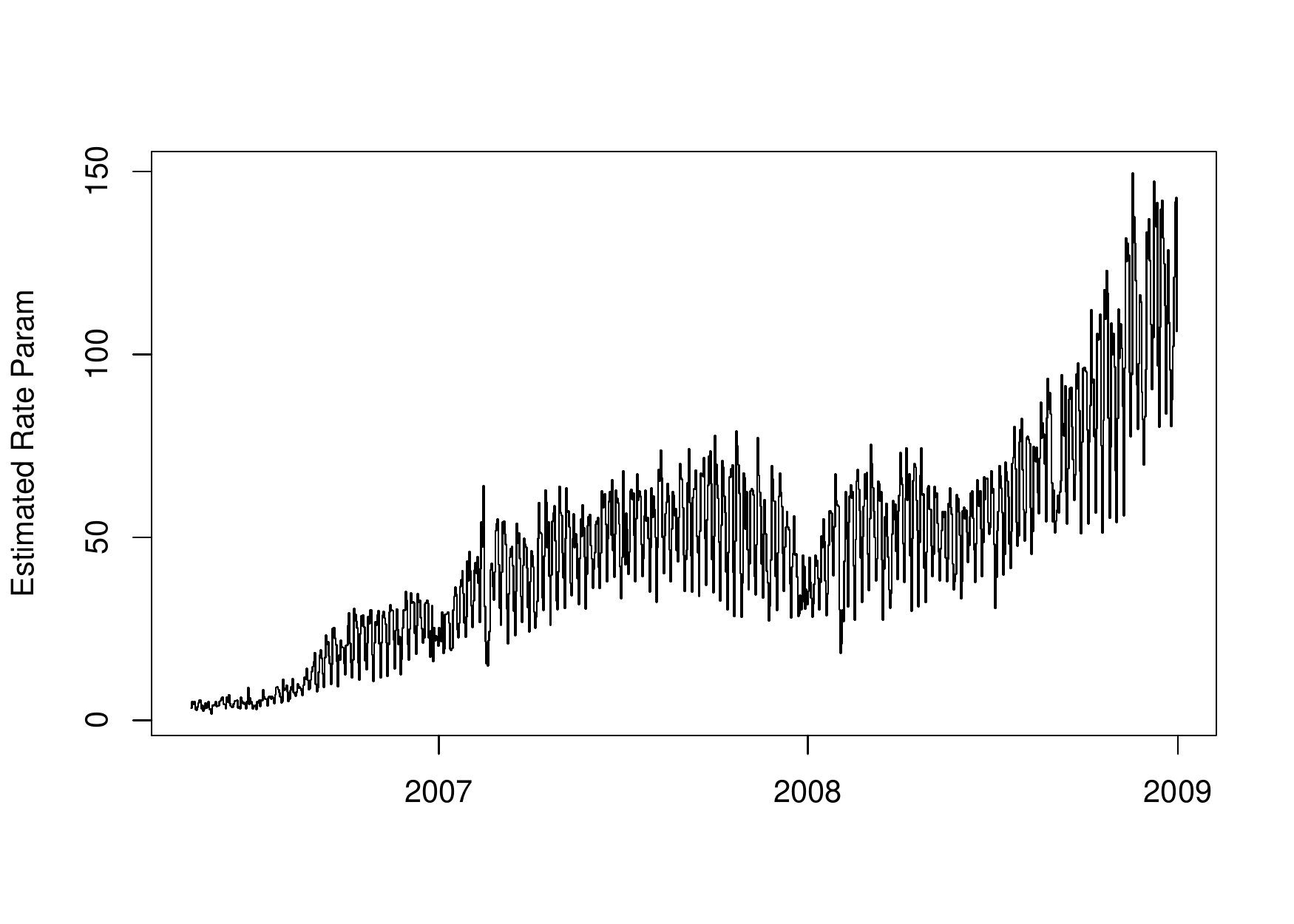}
\caption{MLE's of the rate parameter when a homogeneous Poisson process is fitted to  
 daily cleaned IETs (excluding IETs during 1-8 AM) for the Facebook data from 2006-05-01 to 2008-12-31.}\label{fig:fb_mle}
\end{figure}

Next we exclude timetamps that occur during 1-8 AM on each day {and
obtain reduced data sets for each day.}  Motivated by the observed  decrease in
the burstiness of the process shown in Table~\ref{table:fb_burst}, 
we fit a NHPP to the reduced point process with constant rate across a day but changing rates from day to day.
The estimated rate parameter of the Poisson process is calculated using MLE and this procedure is 
applied to data from 2006-05-01 to 2008-12-31.
Estimation results are given in Figure~\ref{fig:fb_mle}.

We then 
{test each day's  reduced IET data}
against the exponential null.
We also assess our exponential fit by conducting a KS test.
For $\alpha=0.01$, 90.3\% of the time the KS test does not reject the exponential fit.
With $\alpha = 0.05$, the acceptance rate of the KS test is 76.2\%.
Meanwhile, from mid 2007 to mid 2008, the estimated Poisson rate remains relatively stable but keeps increasing 
after mid 2008, which matches what we have observed from Figure~\ref{fig:fb_evo}.

\subsubsection{Summary of Facebook wall posts data}
From the analysis of the Facebook wall posts data, we see that the network growth of the Facebook wall posts (New Orleans) has three different phases.
\begin{enumerate}
\item Prior to the first quarter of year 2006, the point process is
  bursty and we model the network growth by 
a SEPP with an exponential triggering function as specified in \eqref{eq:exp_sepp}. We consider this time period as 
the startup phase of the network.
\item Starting from May 2006, we observe
  a repetitive growth pattern.
  Users are less active on the network during 1-8 AM when
we expect users are asleep. After removing  timestamps occurring between 1--8 AM, the reduced point process becomes less bursty and 
can be modeled by a NHPP where the Poisson rate is constant across a day but changing as we move from day to day.
\item From mid 2007 to mid 2008, the daily estimates of the Poisson
  rate parameter  
come to a relatively stable phase but start increasing again after mid 2008. 
The authors of \cite{viswanath2009evolution} speculate the increase after mid 2008 as a result of the release of Facebook's new design on the wall posts.
\end{enumerate}

\subsection{Dutch Wiki Talk}
We now discuss data from another regional social network, the Dutch
Wikipedia Talk. This dataset has been analyzed in
\cite{wan:wang:davis:resnick:2017} where we fit a linear preferential
attachment model, but here we only consider the timestamp information
in this dataset. We take the cleaned data from
\cite{wan:wang:davis:resnick:2017} where inferred administrative
broadcasts and posts are removed and look at remaining edges created from
2003-01-01 to 2013-12-31. The data forms a directed network where
nodes are users and edges denote user
interactions on the talk pages in Dutch Wikipedia. 

\begin{figure}[h]
\centering
\includegraphics[scale=.5]{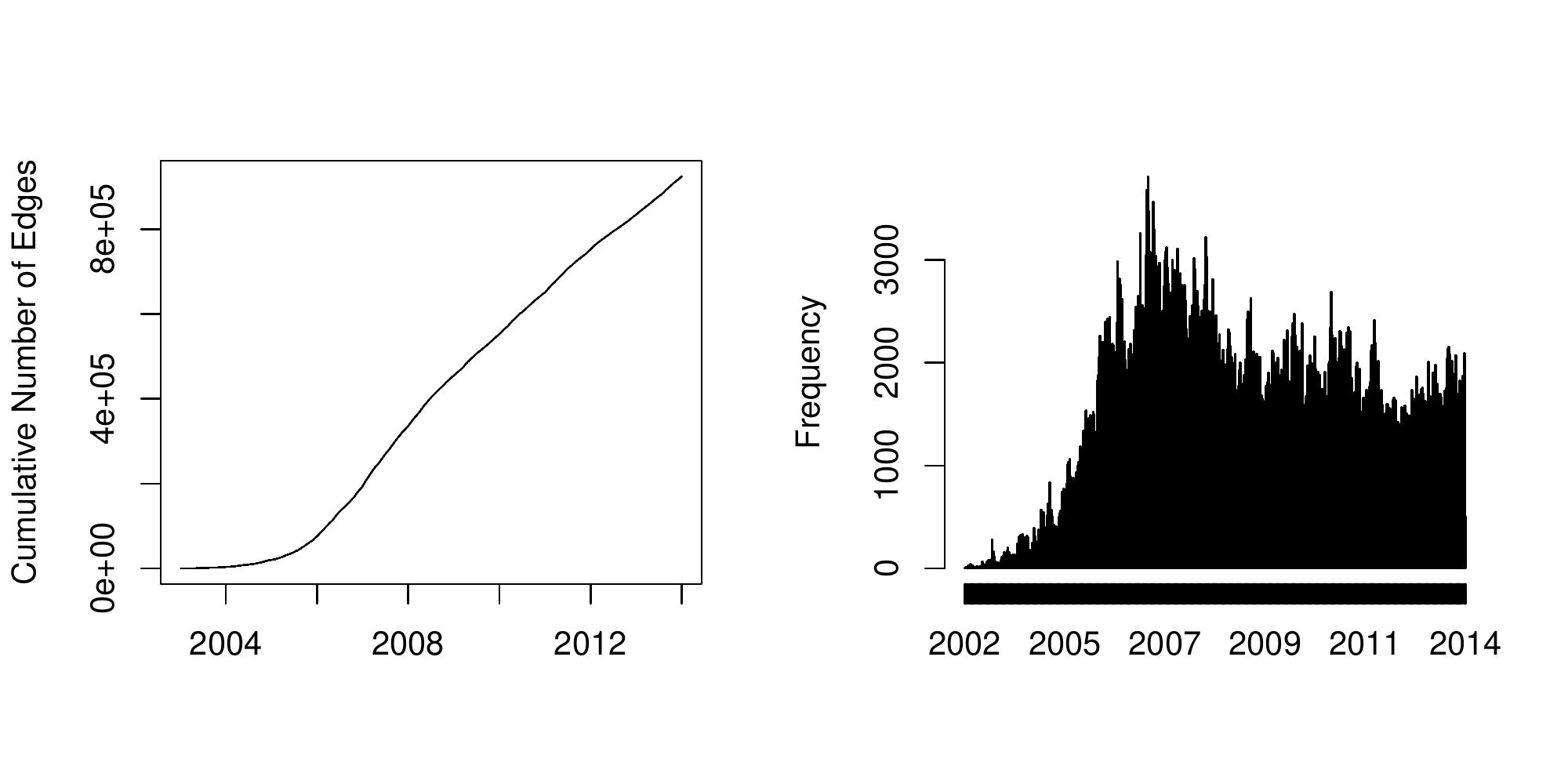}
\caption{The evolution of the Dutch Wiki Talk data. Left: Cumulative
  number of edges over 
time. Right: Total number of new edges added per week.}\label{fig:wiki_nl_evo}
\end{figure}
This dataset contains 924,473 rows and three columns. 
The first two columns represent users' ID and the third column a UNIX timestamp with
the time of a message on one's Wiki talk page.
For each row, the first user writes a message on the talk page of the second user at a timestamp given in the third column.
The evolution of the Dutch Wiki talk network is plotted in Figure~\ref{fig:wiki_nl_evo}. We adjust all timestamps according to 
the Central Europe Time Zone. Different from the Facebook wall post
data, the Dutch Wiki Talk data displays an accelerating growth between
2003 and 2007 but  starting from 2008
the growth pattern is close to linear.

Similar to the Facebook case, after accounting for the poor fit of 
 a SEPP model with a power-law triggering function \eqref{eq:pl_sepp}, we proceed with an exponential triggering function as in \eqref{eq:exp_sepp}.
We first divide the Dutch Wiki talk data from 2003-01-01 to 2005-12-31 into 151 disjoint time windows, each of which contains 500 events,
and fit the SEPP model to each time window using the algorithm
outlined in Section~\ref{subsec:est_sepp} with $p=150$ and
$\Delta=(1/24)/12$. We denote estimated model parameters by
$\widehat{\boldsymbol{\theta}}$ with three components as detailed in
\eqref{eq:nls}. 
With $\widehat{\boldsymbol{\theta}}$, we compute the transformed time $\Lambda(T_i)$, $i=1,2,\ldots,n$ as in \eqref{eq:trans_time} and use a KS
test to examine whether the transformed IETs (calculated using \eqref{eq:trans_time_exp} and \eqref{eq:trans_iet_exp}) are exponentially distributed with rate 1. 
Out of the 151 sets of estimates, the acceptance rate for the KS test is 30.7\% with $\alpha = 0.05$ and 46.7\% with $\alpha = 0.01$.
Also, the ACF plots (not included) do not suggest high autocorrelation among the transformed IETs.

This acceptance rate looks lower than for the Facebook case but
now 500 events are included in each estimation interval as opposed to
200 events for Facebook. With
a larger sample size, we are more likely to have abnormal IETs that
deviate from our model and hence reject the null.
Meanwhile, the Dutch Wiki case also has a similar problem as for
Facebook. At the beginning of the network recording period, 
it takes a long time to accumulate 500 events (10 days or more for the first 22 time windows), which again increases the
possibility of having abnormal observations.
{The bigger sample size, slow network startup and the sensitivity
  of the KS test to
  outliers encourage us to proceed assuming the SEPP model for this
  time period.}

\begin{figure}[h]
\centering
\includegraphics[scale=.6]{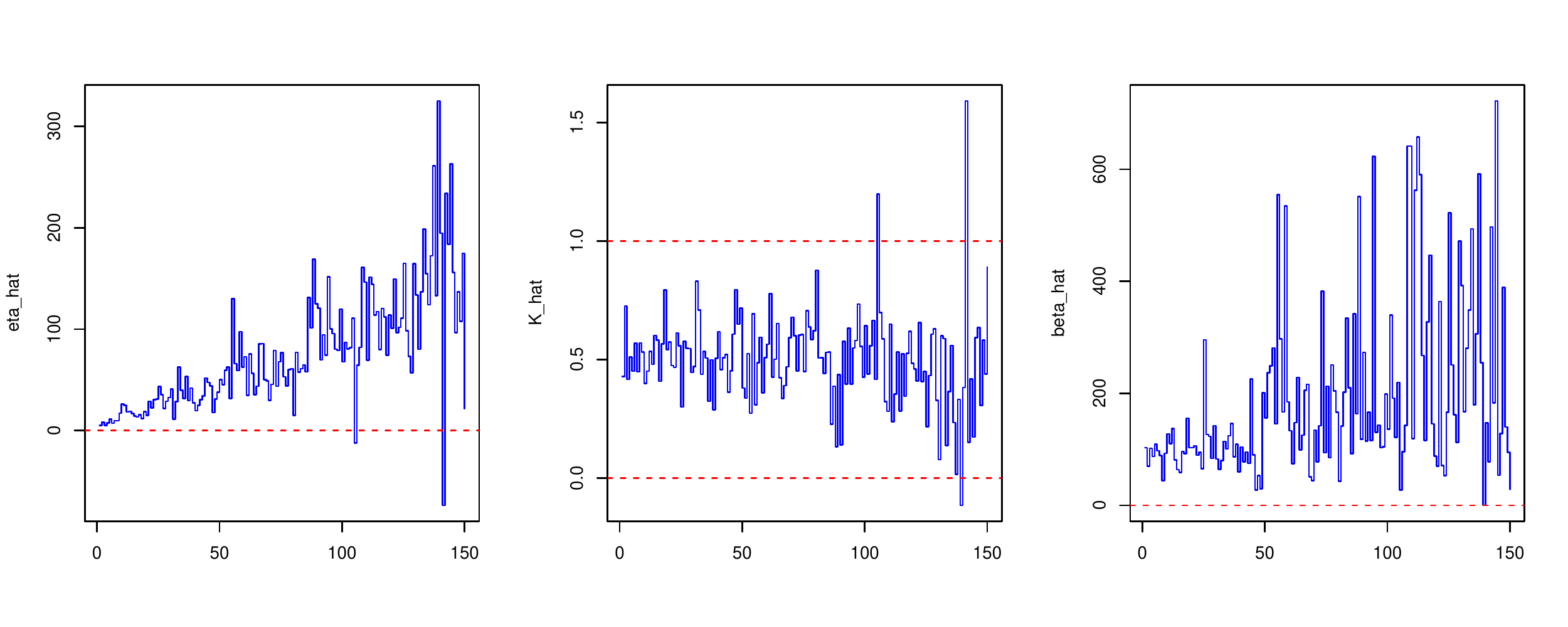}
\caption{Left to right: Estimates for $\eta, K, \beta$ from the
  Dutch Wiki talk data 2003-01-01 to 2005-12-31.  
Each set of estimates are computed using 500 points.
After 45 disjoint time windows, estimates for $\beta$ are more
volatile and
after 100 windows estimates for $K$ occasionally
jump outside of the $(0,1)$ range.}\label{fig:nl_est}
\end{figure}

Estimated values of parameters are given in Figure~\ref{fig:nl_est},
where the left, middle and right panels give $(\widehat{\eta},
\widehat{K}, \widehat{\beta})$, respectively. From the left plot of
Figure~\ref{fig:nl_est}, we observe a {positive} trend in 
$\widehat{\eta}$ with larger variation after 100 time windows
(2005-09-07). Meanwhile, $\widehat{\beta}$ fluctuates more 
after the 46-th time window (2005-02-09) and $\widehat{K}$ jumps
out of the stationary range $(0,1)$ after 100 time windows. 
When $\widehat{K}$ is negative, it is compensated by a large positive
$\widehat{\eta}$. These facts indicate the model {increasingly struggles to fit the data as time progresses.
}

\begin{figure}[h]
\centering
\includegraphics[scale=.6]{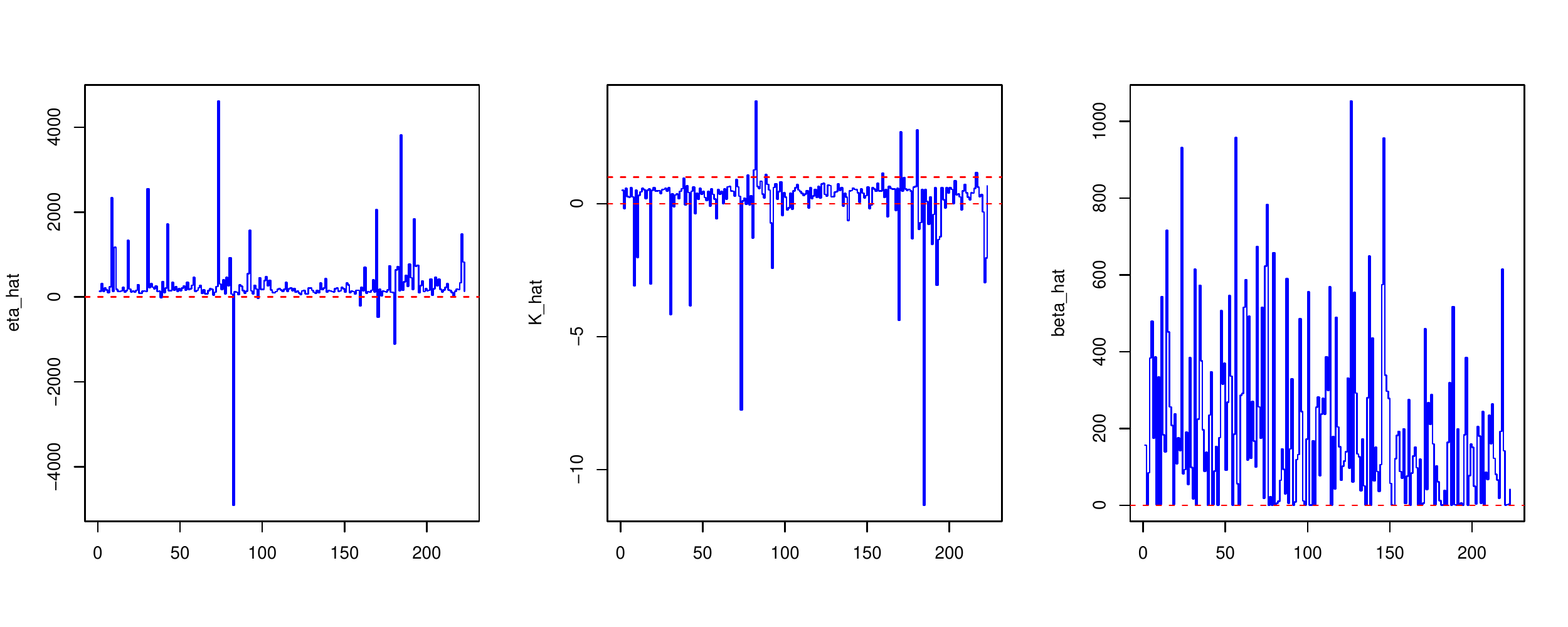}
\caption{Left to right: Estimates for $\eta, K,\beta$ based on the Dutch Wiki talk data from 2006-01-01 to 2006-12-31. 
Each set of estimates are computed using 500 points. 
The three estimated parameters are quite variable.
A large positive $\widehat{\eta}$ is associated with a 
negative $\widehat{K}$,
 which indicates  problematic  fit.
}\label{fig:nl_est_06}
\end{figure}

We further extend the estimation procedure to the data
from 2006-01-01 to 2006-12-31, namely 236 disjoint time windows each
of which has 500 points. The same KS tests are applied to 
transformed IETs and now with $\alpha = 0.05$, the acceptance rate decreases to 19.7\% and 31.4\% when $\alpha = 0.01$. 
Estimated parameter values are plotted in Figure~\ref{fig:nl_est_06}.
We see from Figure~\ref{fig:nl_est_06} that $\widehat{\eta}$ fluctuates more and a large positive $\widehat{\eta}$
is likely to be accompanied by a negative $\widehat{K}$.
Estimated values for $\beta$ are also volatile and sometimes become very close to zero.
 Therefore, we are uncomfortable with the fit and 
seek a different model for data after 2005-02-09. 
 
\begin{figure}[h]
\centering
\includegraphics[scale=.55]{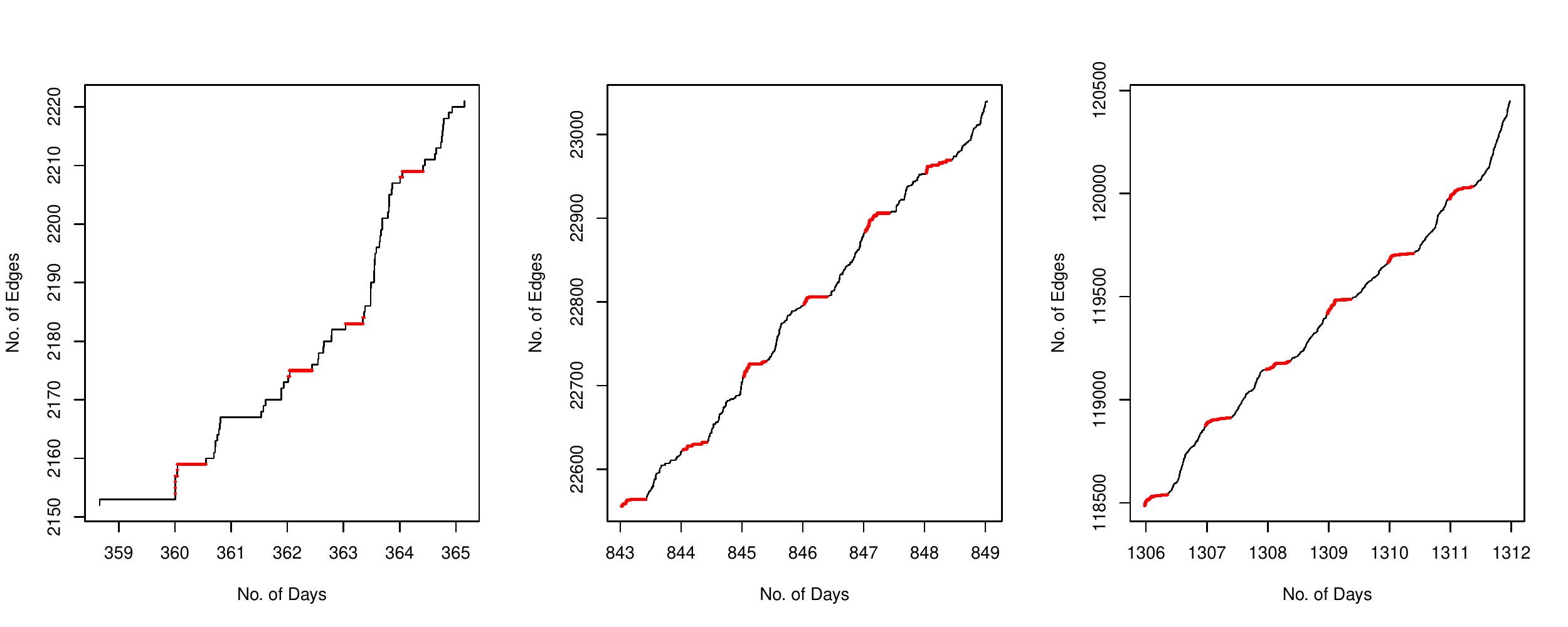}
\caption{Cumulative number of edges during three periods. Left: 2003-10-11 to 2003-10-18.
Middle: 2005-02-07 to 2005-02-13. Right: 2006-05-16 to 2006-05-22.
Red line segments refer to edges created during 0-9 AM each day (Central Europe Time Zone), when we expect sleep interferes with Dutch Wiki talk activity. 
We observe a repeating pattern in the network growth
in the middle and right panels.}\label{fig:nl_sleep}
\end{figure}
Looking at data after 2005-02-09, similar to the Facebook case, we also find a similar repetitive pattern. 
Figure~\ref{fig:nl_sleep} gives an example where we plot the
cumulative number of edges created from
\begin{enumerate}
\item 2003-10-11 to 2003-10-18: the SEPP model fits well.
\item 2005-02-07 to 2005-02-13: the SEPP model starts to fit poorly.
\item 2006-05-16 to 2006-05-22: a {clear} repetitive pattern is observed.
\end{enumerate}
All IETs associated with edges created between 0-9 AM each day are marked as red segments in Figure~\ref{fig:nl_sleep}. 
We see that after the SEPP model starts to return a poor fit {in 2006}, longer
IETs are often associated with the time period when people
usually sleep and are
therefore 
inactive on networks.
\begin{table}[ht]
\begin{center}
\begin{tabular}{cccccc}
\hline
& Yr 2003 & Yr 2004 & Yr 2005 & Yr 2006 & Yr 2007  \\ 
\hline
$B_\text{all}$ & 0.441 & 0.395 & 0.460 & 0.447 & 0.471\\ 
$B_\text{active}$ & 0.448 & 0.401 & 0.375 & 0.257  & 0.213\\ 
\hline
\end{tabular}
\end{center}
\smallskip
\caption{Burstiness parameters for IETs of the Dutch Wiki talk data from 2003 to 2007, where $B_\text{all}$ is computed using all IETs and we calculate $B_\text{active}$
by excluding all IETs occurring during 0-9 AM on each day. Excluding IETs from 0-9 AM decreases the burstiness parameter.}
\label{table:nl_burst}
\end{table}

Table~\ref{table:nl_burst} summarizes the burstiness parameters where
$B_\text{all}$ is computed using all timestamps and we calculate
$B_\text{active}$ 
by excluding timestamps occurring during 0-9 AM on each day. We see
that the point process is {progressively} less bursty after timestamps  
between 0-9 AM are removed.

\begin{figure}[h]
\centering
\includegraphics[scale=.75]{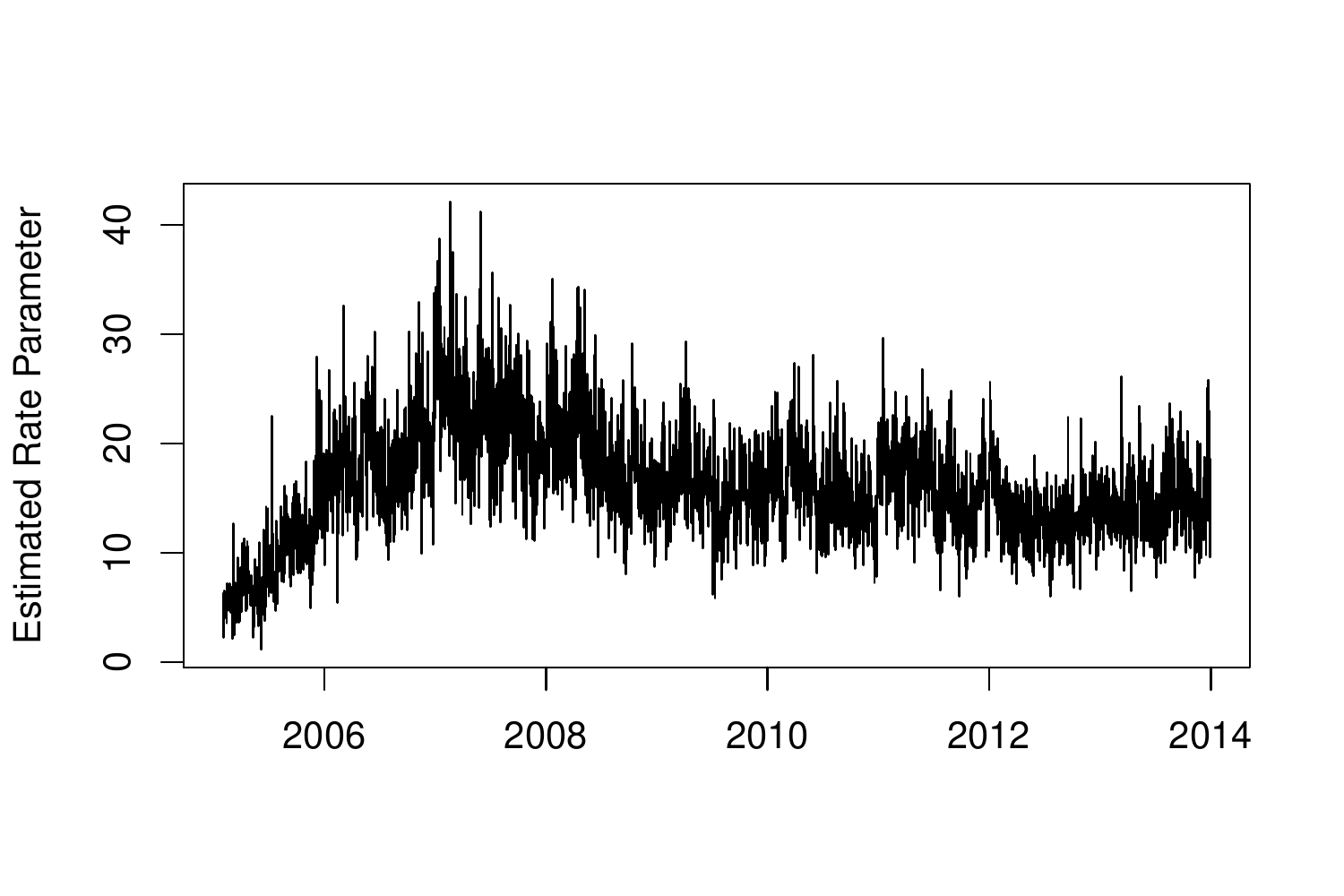}
\caption{MLE's of the exponential rate parameter when a NHPP is fitted to  
 daily cleaned IETs (excluding IETs during 0-9 AM) based
on the Dutch Wiki talk data from 2005-02-01 to 2013-12-31. }\label{fig:nl_rate_daily}
\end{figure}

{Based on the evidence of decreasing burstiness in Table
  \ref{table:nl_burst} and also Figure 
\ref{fig:nl_sleep},}  we exclude timestamps created between
0-9 AM on each day after 
2005-02-01 and fit a NHPP 
to the cleaned data with a constant rate within each day.
We plot estimated daily rate parameters
 in  Figure~\ref{fig:nl_rate_daily} based on
data from 2005-02-01 to 2013-12-31. The Poisson rate estimates are
increasing until 2008 and remain 
relatively stable from 2008 onward.

\begin{table}[ht]
\begin{center}
\begin{tabular}{ccc}
\hline
Year & $\alpha = 0.05$ & $\alpha =0.01$  \\ 
\hline
2005 & 30.9 & 51.5\\
2006 & 22.7 & 40.0\\
2007 & 27.9 & 49.6 \\
2008 & 31.2 & 49.3\\
2009 & 34.5 & 60.0\\
2010 & 36.7 & 60.3\\
2011 & 35.1 & 58.6\\
2012 & 35.1 & 62.7\\
2013 & 32.3 & 55.1\\
\hline
\end{tabular}
\end{center}
\smallskip
\caption{Acceptance rate (\%) of the KS test for a NHPP model fitted to the Dutch Wiki talk data from 2005-02-01 to 2013-12-31.
We assume the rate parameter is constant across the day but varies from day to day.
The fit improves since year 2008.}
\label{table:nl_pois_pass}
\end{table}

{We use a KS test for a null hypothesis of exponentiality applied to
the daily cleaned data.}
We summarize the acceptance rate of the KS test in Table~\ref{table:nl_pois_pass} with $\alpha = 0.01, 0.05$.
We see that the acceptance rate of the KS test is increasing over each year indicating an improvement in the Poisson fit.
Hence, for the Dutch Wiki talk network from 2005 to 2008, we can model
the daily
point process without the points during 0-9 AM
as a NHPP but the estimated rate parameters are in general
increasing. After 2008, the Poisson fit to the cleaned data 
is  good and the estimated rate parameters stay relatively stable.

\subsubsection{Summary of Dutch Wiki talk data}
From the analysis of the Dutch Wiki talk data, we conclude that the growth pattern has three phases:
\begin{enumerate}
\item The point process is bursty up until early 2005 and 
 network growth is   well-modelled
by a SEPP  with an exponential
  triggering function \eqref{eq:exp_sepp}. This is referred to as the
  startup phase of the network.
  
\item From February 2005 onward, we model the growth pattern by a NHPP with constant rate across a day but changing
rates from day to day, plus a daily inactive cycle from 1-9 AM. Noticeably from Figure~\ref{fig:nl_rate_daily}, the daily rates fluctuate more and keep increasing from early 2005 to 2007, which can be
regarded as as the growing phase.
Large fluctuations may be due to abnormal observations in the data, and the large fluctuations may account
for
the lower acceptance rate from the KS test; see
Table~\ref{table:nl_pois_pass}.  
\item Starting from 2008, the Poisson rate estimates (after removing
  timestamps between 1-9 AM) become relatively stable, and we  
{think of} it as the stable phase of the network growth. 
\end{enumerate}

In the interests of parsimony,
one could try to fit an intensity function to the plot in Figure \ref{fig:nl_rate_daily}
of the NHPP model used in the growing and stable phases. For instance, 
by looking for common trends in the rate estimates within a week.
We give a brief discussion later in Section~\ref{subsec:future}.
{However, we have not pursued this carefully.}

\subsection{German Wiki Talk}
In this section, we consider the Wikipedia Talk data in German, which takes a similar format as in the Dutch case, but the dataset
is much larger.
We first clean the data by removing all edges created by ``administrators" and ``bots". This is done by \cite{sun2016wiki}, 
where users of the German Wikipedia are classified 
into three different groups, namely ``normal users", ``administrators" and ``bots". We again look at the data from 2003-01-01 to 2013-12-31.
The cleaned dataset contains 5,863,373 rows and three columns. We
speculate the larger size of German Wikipedia vs Dutch Wikipedia data
is a consequence of the much larger  
number of fluent German speakers worldwide (90--95 million according
to \cite{wikipedia2019german}) compared to Dutch speakers (29 million
by \cite{wikipedia2019dutch}). 

\begin{figure}[h]
\centering
\includegraphics[scale=.6]{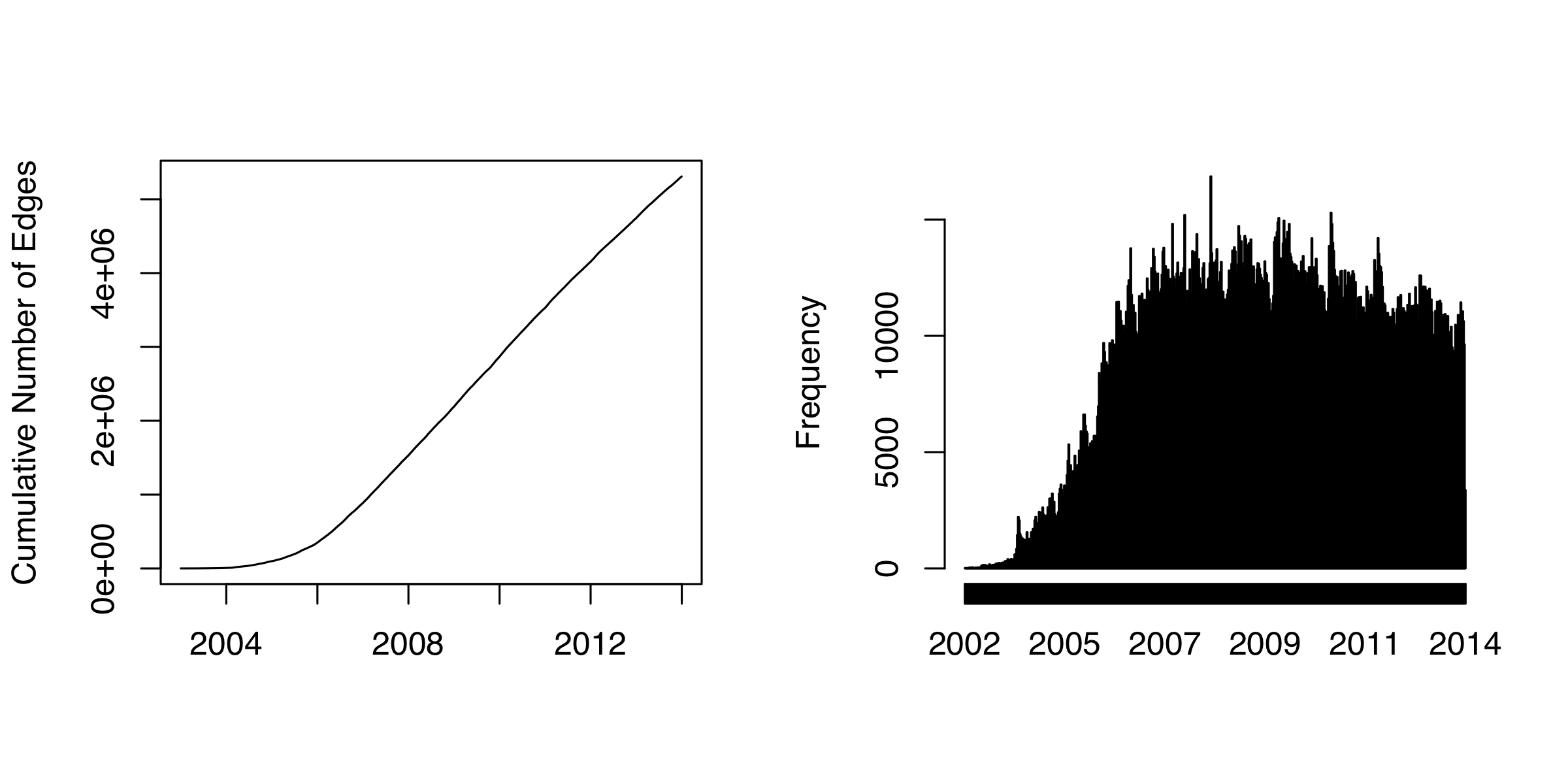}
\caption{The evolution of the German Wikipedia Talk network from 2003-01-01 to 2013-12-31. Left: Cumulative number of edges. Right: The number of edges added per week.}\label{fig:wiki_de_evo}
\end{figure}

Figure~\ref{fig:wiki_de_evo} displays the evolution of this network, where the cumulative number of edges is plotted in the left panel and the
number of new edges added per week is given in the right panel. All timetamps are adjusted to the Central Europe Time Zone.
The shape of the graph looks very similar to the Dutch case, but the German Wiki Talk data grows faster.

\begin{figure}[h]
\centering
\includegraphics[scale=.6]{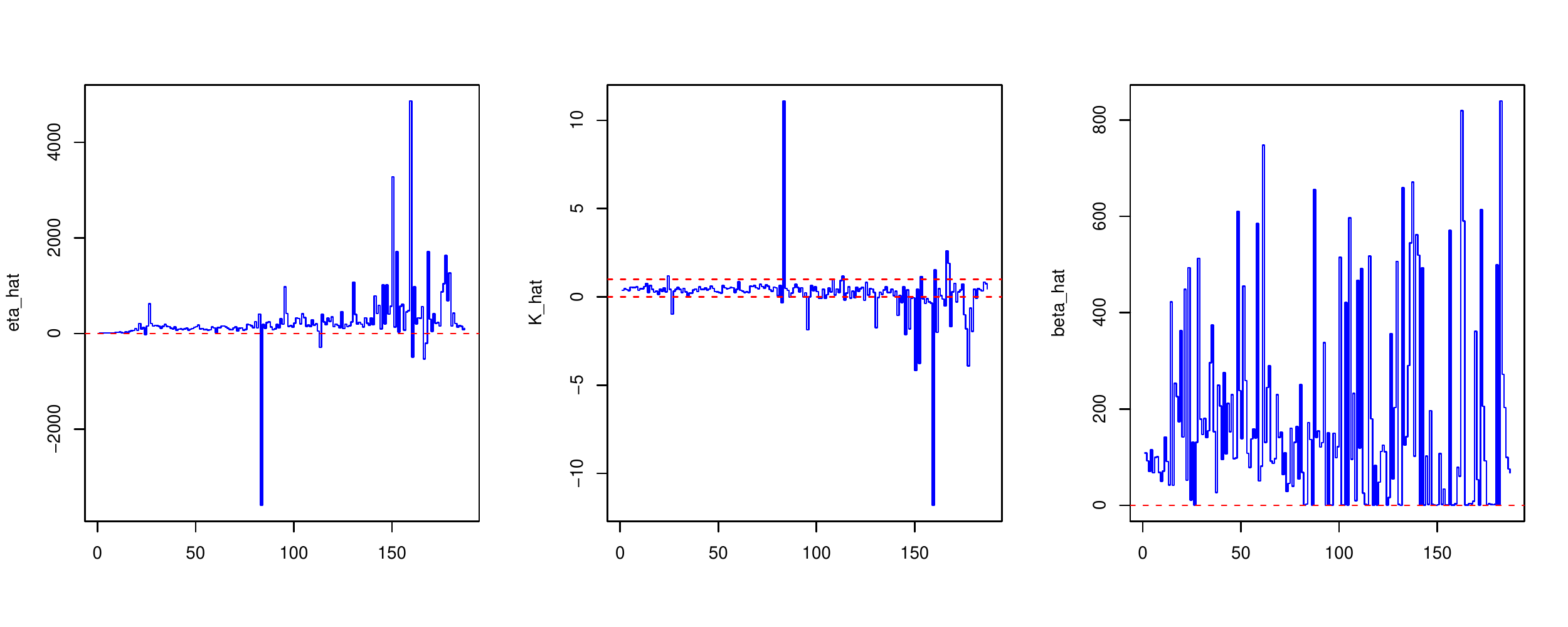}
\caption{Left to right: Estimated values for $\widehat{\eta},
  \widehat{K}, \widehat{\beta}$.
Note that $\widehat{\eta}$ fluctuates more after the 82-nd time window (2004-07-19) and a large $\widehat{\eta}$ 
is offset by a negative $\widehat{K}$, indicating a problematic fit.
}\label{fig:de_est}
\end{figure}

We start by looking at the data from the first two years, namely from
2003-01-01 to 2004-12-31, and divide the data into 196 disjoint time
windows each of which includes 500 points. Fitting a SEPP model with
an exponential triggering function (using the method in
Section~\ref{subsec:est_sepp} with $p=150$ and $\Delta = (1/24)/12$)
to data in these 196 time windows gives estimates of
$\widehat{\boldsymbol{\theta}}$ as in Figure~\ref{fig:de_est}.
Subsequent to the 82-nd time window (2004-07-19), $\widehat{K}$
becomes more volatile and often a negative $\widehat{K}$ is offset by
a large positive value of $\widehat{\eta}$. Fluctuations of the
estimates $\widehat{\beta}$ also increase after the 82-nd time window
and sometimes the estimated value is close to zero.  These
observations from Figure~\ref{fig:de_est} suggest the SEPP model
{struggles to fit the data} after the 82-nd time window.  In fact,
the acceptance rate of the KS test which examines whether the
transformed IETs (calculated using \eqref{eq:trans_time_exp} and
\eqref{eq:trans_iet_exp}) are exponentially distributed with unit mean
is 25.1\% when $\alpha = 0.05$ and 34.2\% when $\alpha = 0.01$, if all
196 time windows are taken into consideration.  However, if only
focusing on the first 81 time windows, the acceptance rate of the same
KS test increases to 36.6\% for $\alpha = 0.05$ and 47.6\% for
$\alpha = 0.01$.  At the beginning of the time record,
longer time periods are
required to accumulate 500 events (10
days or more for the first 16 time windows) and so we are unlikely to obtain
good fits and in light of this observation we conclude that the SEPP
model gives an {acceptable} fit for the network growth until July
2004. We
change the model for data after 2004-07-19.

\begin{figure}[h]
\centering
\includegraphics[scale=.6]{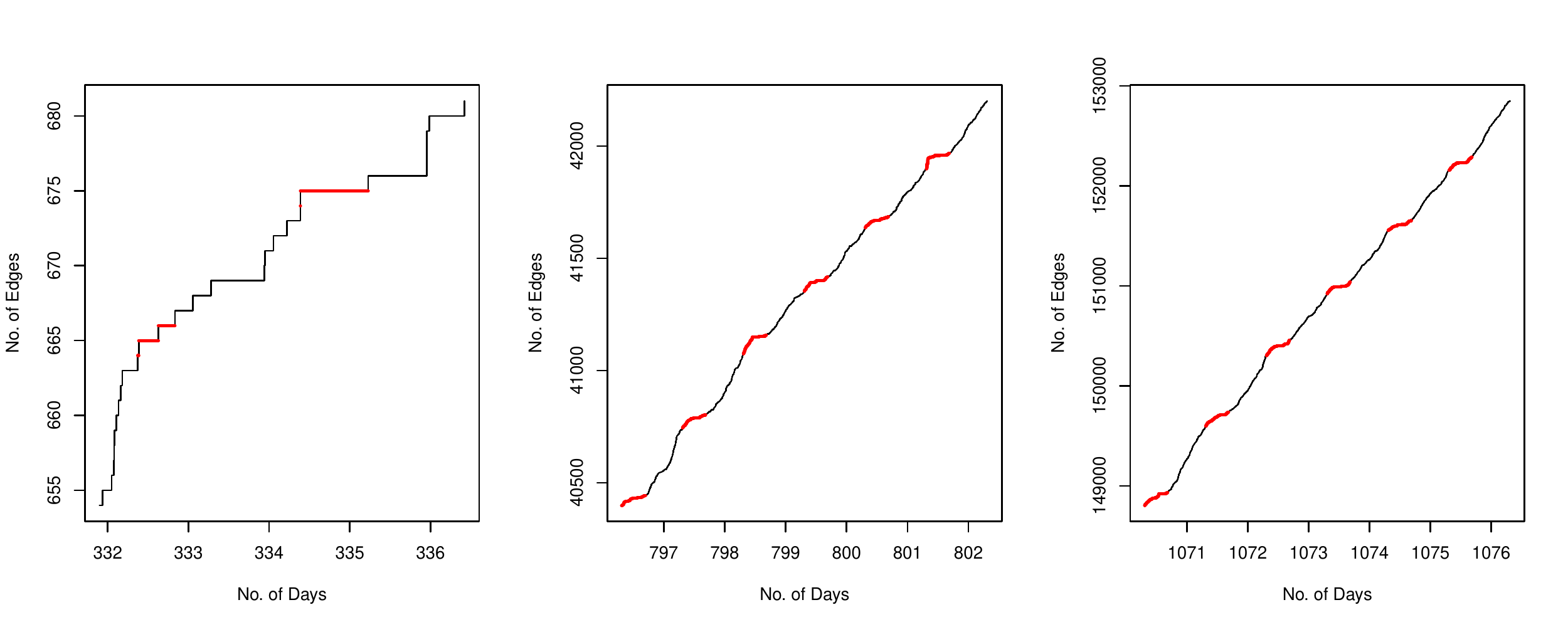}
\caption{Cumulative number of edges during three periods. Left: 2003-04-06 to 2003-04-12.
Middle: 2004-07-15 to 2004-07-21. Right: 2005-04-15 to 2005-04-21.
Red line segments give edges created during 0-9 AM each day (Central Europe Time Zone), 
when  sleep interferes with German Wiki talk activity. 
Network growth has a repeating pattern  in the
 middle and right panels.} 
\label{fig:de_sleep}
\end{figure}
Similar to the Dutch Wiki case, a repeating pattern is observed in the growth
of the German Wiki talk network. Figure~\ref{fig:de_sleep} plots the cumulative number of edges during three time periods:
\begin{enumerate}
\item 2003-04-06 to 2003-04-12: the SEPP model fits well.
\item 2004-07-15 to 2004-07-21: the SEPP model fits poorly, suggesting changes in modeling.
\item 2005-04-15 to 2005-04-21: the point process exhibits a repetitive pattern.
\end{enumerate}
Red line segments correspond to edges created during 0-9 AM on each day. From the middle and right panels of 
Figure~\ref{fig:de_sleep}, as expected, we see a repeating pattern and users are less active from 0 to 9 AM on each day.
This agrees with the findings in the Dutch Wiki case.

\begin{table}[ht]
\begin{center}
\begin{tabular}{ccc}
\hline
Year & $\alpha = 0.05$ & $\alpha =0.01$  \\ 
\hline
2004 & 21.1 & 33.7\\
2005 & 51.5 & 73.4\\
2006 & 41.1 & 63.0\\
2007 & 20.2 & 41.1 \\
2008 & 26.3 & 50.4\\
2009 & 27.1 & 49.3\\
2010 & 27.4 & 48.8\\
2011 & 44.7 & 67.9\\
2012 & 40.8 & 66.3\\
2013 & 41.1 & 59.2\\
\hline
\end{tabular}
\end{center}
\smallskip
\caption{Acceptance rates (\%) of the KS test for a NHPP model fitted to the German Wiki talk data from 2004-07-01 to 2013-12-31.
The fit improves subsequent to year 2005. On average, the acceptance rate is 35.9\% when $\alpha = 0.05$ and 58.2\% when $\alpha = 0.01$.}
\label{table:de_pois_pass}
\end{table}

\begin{figure}
\centering
\includegraphics[scale=.6]{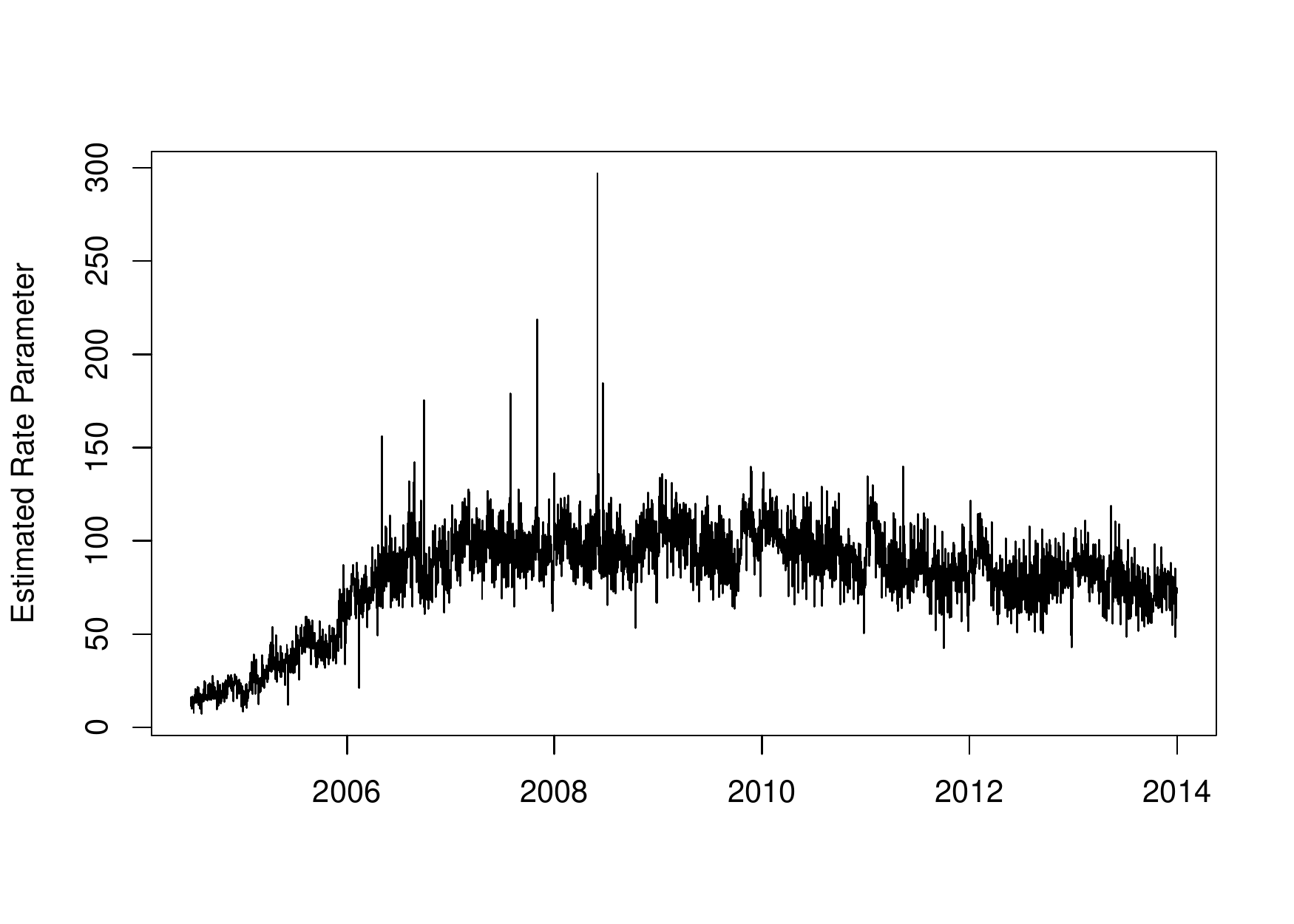}
\caption{MLE's of the rate parameter when a NHPP is fitted to  
 daily cleaned IETs (excluding IETs during 0-9 AM) based
on the German Wiki talk data from 2004-07-01 to 2013-12-31.}\label{fig:de_pois_rate}
\end{figure}

For the data from 2004-07-01 to 2013-12-31,
we remove all timestamps during 0-9 AM and fit a
NHPP model to the cleaned data such that the rate parameters are
constant across the day but vary from day to day. The acceptance rate
for the KS test assessing the Poisson fit is reported in
Table~\ref{table:de_pois_pass}.  In general, fits after 2005 are
better then before
and on average, the acceptance rate is 35.9\% when
$\alpha = 0.05$ and 58.2\% when $\alpha = 0.01$.  The estimated rate
parameters (using reduced daily data) are plotted in
Figure~\ref{fig:de_pois_rate}. The rate estimates {exhibit
increasing trend until roughly}
2007 and stay relatively stable afterwards.  Estimates fluctuate more
during 2007-2010, which is associated with lower acceptance rates of
KS test in Table~\ref{table:de_pois_pass}. This may be due to
abnormal IETs during that particular time period which cannot be
caught by the Poisson model.

\subsubsection{Summary of German Wiki talk data}
The conclusion for the German Wiki data is similar to the Dutch case.
The network is in startup phase until July 2004 during which the
growth can be modeled by a SEPP with an exponential triggering
function.  After the startup phase ends, edge creation
can be viewed as a NHPP with constant rate within a day
but different rates from day to day, plus a daily inactivity time
period during which users are presumably asleep.

The difference between the German Wiki data and the Dutch one is the
growth rate of the network.  Although both were founded in 2001
(\cite{wikipedia:nl, wikipedia:de}),  German Wikipedia has more
articles and the total number of German speakers (90--95 million)
worldwide is also much larger than that of Dutch speakers (29 million)
(\cite{wikipedia2019dutch, wikipedia2019german}). These may accelerate
the growth of the network and shrink the length of the startup
phase.

\subsection{Math Overflow}
After seeing three regional communication networks that are
geographically localized, for contrast we investigate another network
which is not regional. The MathOverflow data includes interactions
from the ``StackExchange" site ``MathOverflow".  The network was
launched September 28, 2009. Types of interactions on MathOverflow
include answering a question from another user, commenting on another
user's question and commenting on another user's answer.  In the
graphical representation of this network, users are  nodes
and online interaction 
between two users is regarded as a directed edge.  The dataset
contains 506,550 edges (rows) and covers the time period from
2009-09-28 to 2016-03-06.  There are three columns in the dataset.
The first two columns represent users' ID and the third column is a
UNIX timestamp with the time at which an interaction is made.  For
each row, the first user interacts with the second user in one of the
three ways given above. 

\begin{figure}[h]
\centering
\includegraphics[scale=.6]{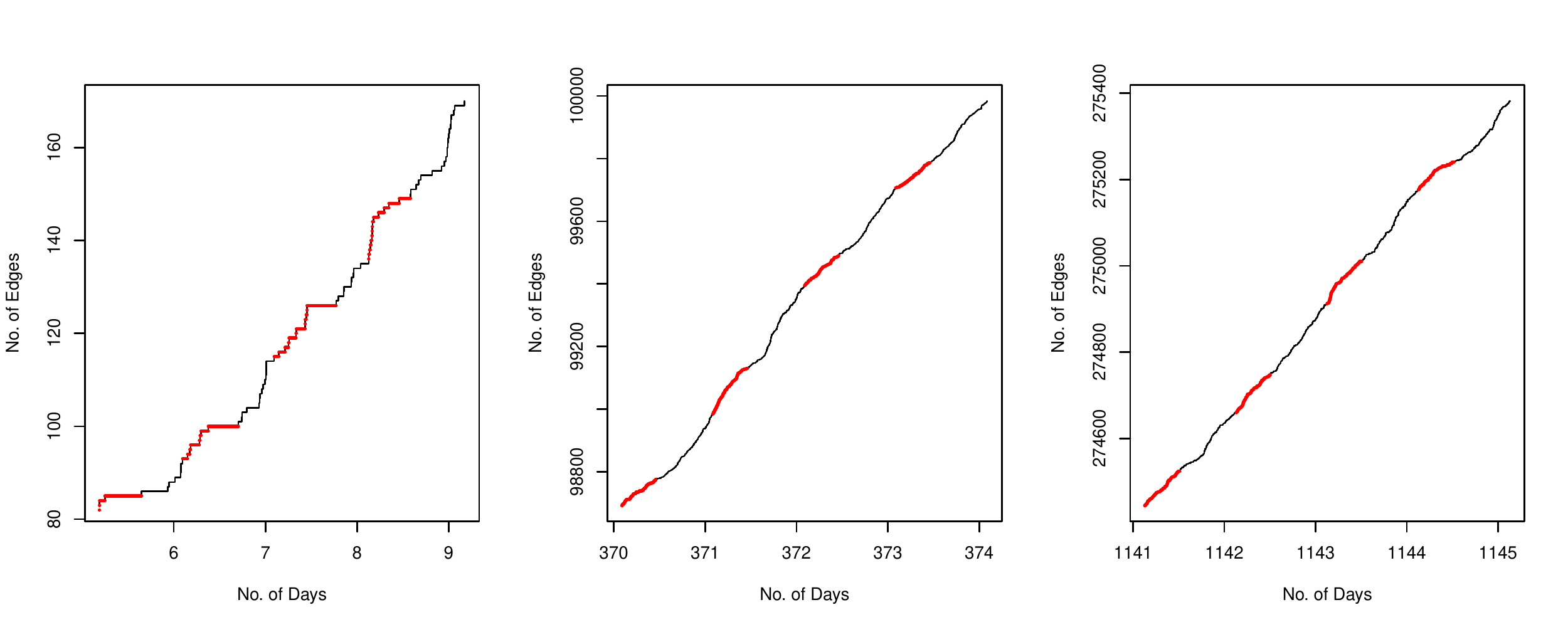}
\caption{Cumulative number of edges from the MathOverflow data during
  three representative time periods: (1) 2009-10-03 -- 2009-10-07
  (left; the beginning of the network); 
(2) 2010-03-02 -- 2010-03-06 (middle; about one year after the launch of MathOverflow); (3)
2012-11-12 -- 2012-11-16 (right; about two years after the launch of MathOverflow). 
Red line segments refer to edges created during 0-9 AM each day, based on
US Central Time.
Unlike the regional networks, the repeating pattern with a nightly inactivity period is not observed here, perhaps because
the MathOverflow website is used by users worldwide and thus is not regional.}\label{fig:math_sleep}
\end{figure}

Since the MathOverflow website can be accessed by users worldwide, there is no obvious choice of a particular time zone to adjust
the timestamps associated with the creation of each edge. We simply set the time zone to be the US Central Time and Figure~\ref{fig:math_sleep}
plots the cumulative number of edges within three time periods:
\begin{enumerate}
\item 2009-10-03 -- 2009-10-07, which is the very beginning of the network.
\item 2010-10-03 -- 2010-10-07, approximately one year after the network has been launched.
\item 2012-11-12 -- 2012-11-16, approximately two years after the network has been launched.
\end{enumerate}
Comparing Figure~\ref{fig:math_sleep} with Figure~\ref{fig:rep_fb}, \ref{fig:nl_sleep} and \ref{fig:de_sleep}, we see that
the daily repeating pattern vanishes as the network evolves and no nightly inactivity period has been observed.

\begin{figure}[h]
\centering
\includegraphics[scale=.6]{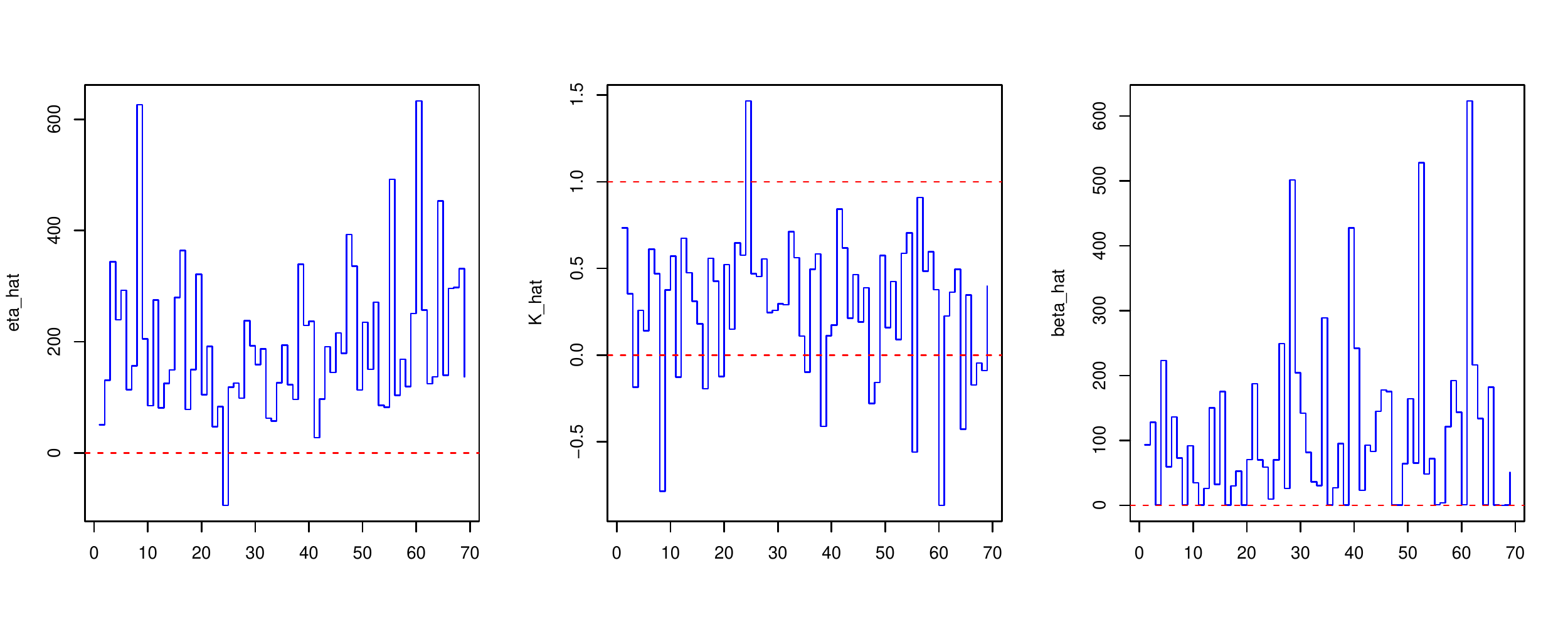}
\caption{Estimated values for $(\widehat{\eta}, \widehat{\beta})$ (left) and $\widehat{K}$ (right) based on 
the MathOverflow data from 2009-10-01 to 2010-02-28. Each set of estimates is computed using 500 points.
Note that 15 out of 69 estimated $\widehat{K}$ values are negative.}
\label{fig:math_est}
\end{figure}

We try fitting a SEPP model to the first 5 months of the network data (from 2009-10-01 to 2010-02-28) by creating 69 consecutive
 time windows each of which contains 500 points. Estimation results are presented in Figure~\ref{fig:math_est}.
Then we again use a KS test to compare the transformed IETs (calculated using \eqref{eq:trans_time_exp} and \eqref{eq:trans_iet_exp}) with the null hypothesis of an exponential distribution with unit mean.
Although the acceptance rate of the KS test is 42.0\% when $\alpha=0.05$ and 58.0\% when $\alpha = 0.01$, 
15 out of the 69 $\widehat{K}$ values
(the middle panel of Figure~\ref{fig:math_est})
are negative. 
When inspecting the ACF plots for transformed IETs (not included here), we see large autocorrelations, which 
raises the concern on the goodness of fit for the SEPP model.
So here even though the KS test has a reasonable acceptance rate, we think the SEPP is not suitable to model the early-stage
growth of the MathOverflow.

\begin{table}[ht]
\begin{center}
\begin{tabular}{ccc}
\hline
Year & $\alpha = 0.05$ & $\alpha =0.01$  \\ 
\hline
2010 & 67.1 & 84.9\\
2011 & 67.7 & 88.2\\
2012 & 76.4 & 92.3\\
2013 & 70.7 & 86.3 \\
2014 & 66.3 & 85.2\\
2015 & 65.5 & 87.9\\
\hline
\end{tabular}
\end{center}
\smallskip
\caption{Acceptance rate (\%) of the KS test for a NHPP model fitted to the MathOverflow data from 2010-01-01 to 2015-12-31.
The acceptance rate is high, but autocorrelation between IETs is not negligible.}
\label{table:math_pois_pass}
\end{table}

Given that no clear repetitive daily inactivity period has been observed, we try fitting a NHPP
to the timestamps such that IET data within one day is tested against an exponential null with a constant rate, but the rate parameter varies from day to day. 
The rate parameter
of the Poisson process is estimated using MLE. The acceptance rates of the KS test for the MathOverflow data from 2010-01-01 to 2015-12-31 
are summarized in Table~\ref{table:math_pois_pass}. The acceptance rates stay very high for each year but at the same time the
autocorrelation among IETs (ACF plots not included here) is also significant, indicating more analysis could be conducted to try to account for the correlation.

Therefore, we conclude that the SEPP $+$ NHPP model does not provide a good fit for the MathOverflow data and the growth pattern for this network which is not geographically concentrated is different from 
the three regional networks we looked at.

\section{Discussion}\label{sec:discuss}
\subsection{Common growth pattern}
In this paper, we study four social network datasets, three of which are 
geologically concentrated, namely
Facebook wall posts for users in New Orleans, Dutch and German Wikipedia talk messages.
The MathOverflow data does not have the geological concentration property.
After comparing these four datasets, we see that regional social networks display some common evolution pattern:
\begin{enumerate}
\item Startup phase: When the network is building up, the network evolution can be modeled by a SEPP with an exponential triggering function.
During this period, longer IETs are not necessarily associated with expected daily inactive periods for users.
Meanwhile, the length of the startup phase varies and may depend on the number of users of the network.
For instance, in the German Wiki talk case, the startup phase disappears within roughly 1.5 years after which the SEPP model
starts to fit poorly. In contrast, the startup phase in the Dutch Wiki case lasts for more than two years.
The different startup lengths may be a consequence of the
difference in the number of possible users of the network. 
According to Wikipedia \cite{wikipedia2019dutch, wikipedia2019german}, 
the total number of German speakers worldwide is about 90--95 million whereas that of Dutch
speakers in the world is around 29 million. More available users have the potential to accelerate the growth of the network, making 
the startup phase disappear more quickly.
\medskip

\item Growing phase: After the network has attracted some attention, it enters a growing phase and the daily evolution can be 
modeled by a NHPP, where the Poisson rate changes but stays constant across each day, plus a nightly inactive period. 
The estimated rate parameters of the Poisson process while not monotone are trending upwards during
the growing phase.
\medskip

\item Stable phase: The regional network becomes relatively saturated and the the overnight inactivity persists.
After reaching the stable phase, the rate parameters from the Poisson process remain stable unless some new external stimulus is brought to the network.
For example, in the Facebook case, we observe a sudden increase in the
estimated rate parameters since mid-2008. This may be due to 
speculated in Facebook's release of a
new site design on July 20, 2008  \cite{viswanath2009evolution} .
\end{enumerate}

This three-stage growth pattern does not apply to networks that are
not geographically  concentrated, e.g. the MathOverflow data.
Although the acceptance rate of a Poisson model is high, the serial
dependence among IETs is  not negligible {and should be accounted for.}

\subsection{Future direction}\label{subsec:future}
This study gives rise to open questions for future research.
\begin{itemize}
\item How can we  model the three-phase network
  evolution theoretically?  Interactions among
  individual users in a network are sometimes modeled by a multivariate Hawkes
  process, {where users are}  influenced by neighbors, authorities
  and communities (cf. \cite{zhou2013learning,farajtabar2014shaping}).
  However, considering the whole network evolution, the transition
  from SEPP to NHPP has not been carefully analyzed. From simulation
  results, simply superposing SEPPs does not always give a process
  that is close to Poisson, and the transition seems to depend on the
  choice of parameters in the SEPP.

\item How can we find a more parsimonious model for regional networks
  in growing or stable phases?  One possible way is to investigate any
  common trend within a week. For example, Figure~\ref{fig:de_wk_pois}
  collects the MLE's of daily Poisson rates in 2012 based on the
  German Wiki data and groups the estimates by week. Labels 1--7 on
  the x-axis corresponds to Monday -- Sunday in a week,
  respectively. The red solid line highlights the average rate
  estimates on each day of the week.  A decreasing trend is observed,
  suggesting that people are less active on the German Wiki talk pages
  during weekends. Hence, one may consider smoothing the daily Poisson
  rate estimates by taking such weekly pattern into account.
\begin{figure}[h]
\centering
\includegraphics[scale=.6]{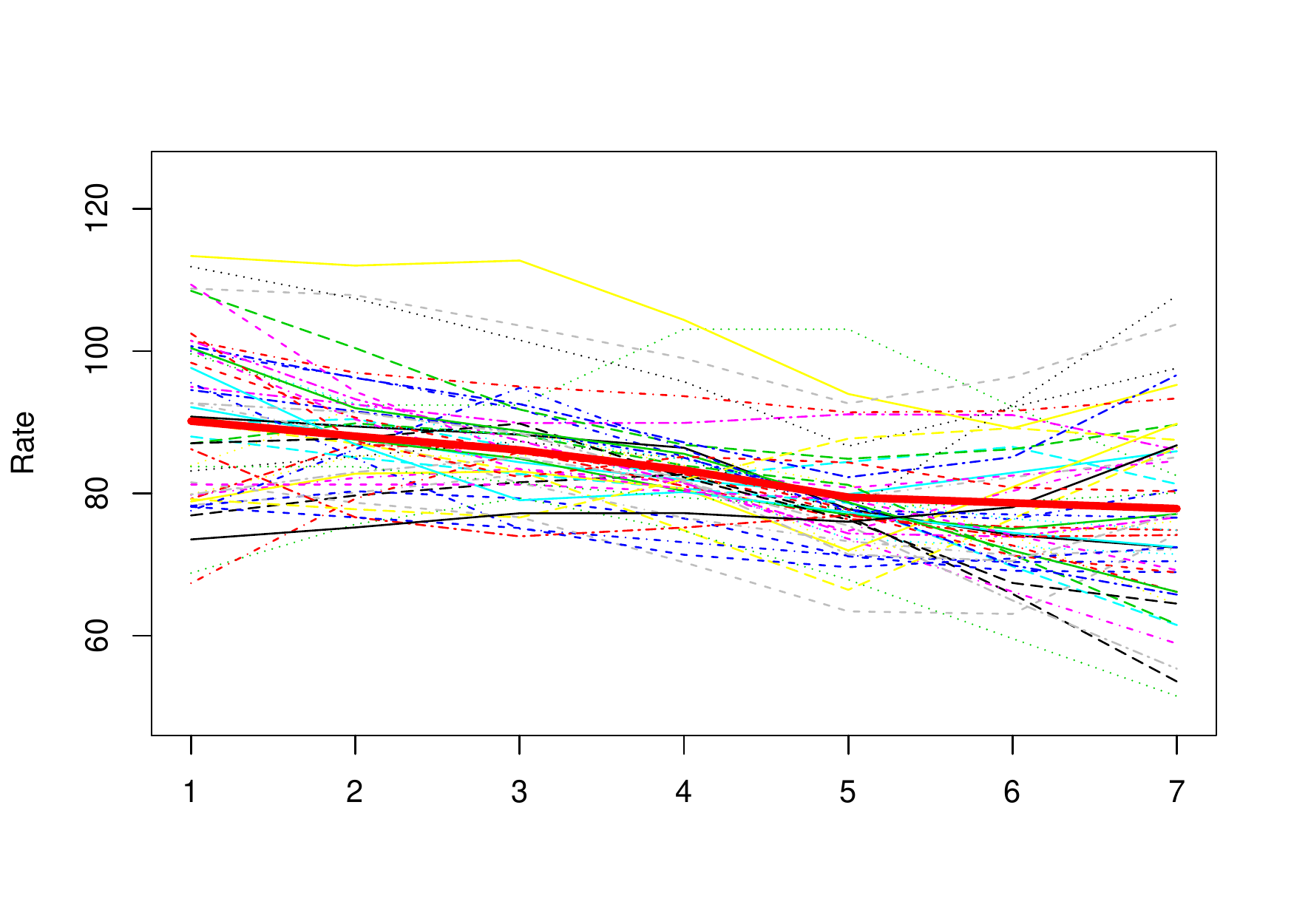}
\caption{MLE's of daily Poisson rates for the German Wiki data in 2012. Estimates are grouped by week so 1--7 labels on the 
x-axis represents Monday -- Sunday in a week, respectively. The red solid line marks the average rates on each day of the week.
We see a decreasing trend over the week, indicating that
users are less active during weekends.}\label{fig:de_wk_pois}
\end{figure}
\end{itemize}

 \bibliography{./bibfile}
 \end{document}